\documentclass[a4paper,12pt]{article}
\usepackage{graphicx} 
\usepackage{dcolumn}
\usepackage{graphicx}
\usepackage{amsmath}
\usepackage{amsfonts}
\usepackage{amssymb}
\usepackage{psfrag}
\usepackage{wrapfig}
\usepackage{subfigure}
\usepackage{makeidx}
\usepackage{bm}
\usepackage{epsf}
\usepackage{hyperref}
\usepackage{color}
\usepackage{cases}
\usepackage{multirow}
\usepackage{booktabs} 
\usepackage{makecell} 
\usepackage{float}

\newcommand{\be}{\begin{equation}}
\newcommand{\ee}{\end{equation}}
\newcommand{\bea}{\setlength\arraycolsep{2pt} \begin{eqnarray}}
\newcommand{\eea}{\end{eqnarray}}
\newcommand{\nn}{\nonumber}

\usepackage{authblk}

\usepackage{color}
\usepackage[normalem]{ulem}
\usepackage[dvipsnames]{xcolor}
\usepackage{cite}

\def\ft#1#2{{\textstyle{\frac{\scriptstyle #1}{\scriptstyle #2} } }}

\def\0{{\sst{(0)}}}
\def\1{{\sst{(1)}}}
\def\2{{\sst{(2)}}}
\def\3{{\sst{(3)}}}
\def\4{{\sst{(4)}}}
\def\5{{\sst{(5)}}}
\def\6{{\sst{(6)}}}
\def\7{{\sst{(7)}}}
\def\8{{\sst{(8)}}}
\def\sst#1{{\scriptscriptstyle #1}}

\title{\vspace{-1.5cm}Light rings and optical appearances of naked singularities, solitons, and black holes in beyond Horndeski gravity}

\author[1,2]{Hyat Huang\thanks{\href{mailto:hyat@mail.bnu.edu.cn}{hyat@mail.bnu.edu.cn}}}
\author[3]{Jutta Kunz\thanks{\href{mailto:jutta.kunz@uni-oldenburg.de}{jutta.kunz@uni-oldenburg.de}}} 
\author[4]{Rashmi Uniyal\thanks{\href{mailto:runiyal687@gmail.com}{runiyal687@gmail.com}}}
\author[1]{Xiao Qian Wang}

\affil[1]{School of Physics, Jiangxi Normal University, Nanchang 330022, China}
\affil[2]{Jiangxi Provincial Key Laboratory of Advanced Electronic Materials and Devices, Jiangxi Normal University, Nanchang 330022, China}
\affil[3]{Institut f\"ur  Physik, Universit\"at Oldenburg, Postfach 2503,
D-26111 Oldenburg, Germany}
\affil[4]{Department of Physics, Rajendra Singh Rawat Government Degree College, Rajgarhi Road, Barkot 249141, Uttarakhand, India}

\date{\today}

\begin{document}

\maketitle

\begin{abstract}

We investigate the geodesic structure and optical appearance of compact objects with primary scalar hair in shift- and parity-symmetric beyond Horndeski gravity. 
The analytic solution considered here depends on a theory parameter and a dimensionless mass parameter \cite{Bakopoulos:2023sdm}.
For a fixed theory parameter, varying the mass traces a family of static spacetimes that can interpolate between timelike naked singularities, regular solitons, regular black holes, Reissner-Nordström-like black holes, multi-horizon black holes, and Schwarzschild-like black holes. 
We classify these branches by their horizon structure and analyze null and timelike geodesics, focusing on light rings, innermost stable circular orbits, and static spheres. 
We then compute thin-disk optical images by ray tracing. 
We find that the number of horizons is not directly encoded in the image: 
horizonless objects can show shadow-like central depressions, while multi-horizon black holes can closely resemble single-horizon black holes when their exterior light ring and disk structures are similar. 
Thus, the optical appearance is governed mainly by the 
photon potential and the disk inner edge, with the deeper horizon structure leaving only an indirect imprint.
Quantitative radial-profile diagnostics confirm that the degeneracy is mainly morphological: the profiles differ at fixed impact parameter, but become much closer after rescaling by the critical impact parameter.
These results provide a concrete example of how distinct compact object branches in beyond Horndeski gravity can share similar observational signatures.

\end{abstract}

\newpage

\section{Introduction}

Compact objects beyond the Kerr and Schwarzschild paradigm provide an important arena for testing gravity in the strong-field regime \cite{Will:2018bme,Faraoni:2010pgm,Berti:2015itd,CANTATA:2021ktz}.
In many modified theories of gravity, the space of static, spherically symmetric solutions is richer than in general relativity and may contain black holes with additional fields, horizonless compact objects, regular geometries, or naked singularities. 
Understanding how these different configurations are organized within a given theory is essential both for clarifying their theoretical properties and for identifying possible observational signatures \cite{LIGOScientific:2016aoc,LIGOScientific:2016sjg,LIGOScientific:2018mvr,LIGOScientific:2020ibl,EventHorizonTelescope:2019dse,EventHorizonTelescope:2019ggy,EventHorizonTelescope:2022wkp,EventHorizonTelescope:2022exc}.

The increasing angular resolution of horizon-scale observations has made the optical appearance of compact objects a useful probe of strong-field gravity. 
Images of accreting black holes, such as those obtained by the Event Horizon Telescope, are mainly sensitive to the exterior photon propagation and to the emission region around the compact object. 
It is therefore important to understand which image features are robust indicators of horizons or light rings, and which can also arise from horizonless compact objects or from the assumed structure of the emitting disk.

A particularly interesting class of examples arises in shift- and parity-symmetric beyond Horndeski theories \cite{Gleyzes:2014dya,Gleyzes:2014qga,Kobayashi:2019hrl}.
These theories belong to the broader family of degenerate higher-order scalar-tensor theories, which extend the Horndeski framework \cite{Horndeski:1974wa}.
Their degeneracy prevents the propagation of an additional Ostrogradsky degree of freedom, while still allowing nontrivial derivative interactions of the scalar field \cite{Langlois:2015cwa,BenAchour:2016fzp,Langlois:2017mdk}.
Recently, analytic compact object solutions with \emph{primary scalar hair} were constructed in this framework \cite{Bakopoulos:2023fmv,Bakopoulos:2023sdm}. 
Unlike secondary hair, primary hair is not fixed by the conserved charges of the spacetime, but instead associated with nontrivial interactions in the Lagrangian.
Therefore, it can lead to families of solutions with properties that differ qualitatively from the Schwarzschild case.

These families of analytic compact object solutions are controlled by coefficients $\{c_n\}$ in the scalar-tensor Lagrangian, where each coefficient $c_n$ multiplies a term proportional to $X^{n/s}$, with $X$ denoting the scalar kinetic density. 
Thus, the ratio $n/s$ labels the order of the corresponding contribution in the kinetic expansion and distinguishes different members of the analytic solution family.
This provides a useful setting in which the role of primary hair can be studied systematically. 
The lowest nontrivial member of this family has $n/s=2$ and corresponds to the black-hole solution with primary scalar hair found by Bakopoulos, Charmousis, Kanti, Lecoeur and Nakas \cite{Bakopoulos:2023fmv}.
Its thermodynamic, dynamical and optical properties have been investigated subsequently \cite{Erices:2024lci}.

In the present work we consider the next member of this family, constructed recently by Bakopoulos, Chatzifotis and Nakas \cite{Bakopoulos:2023sdm}. 
This family has a considerably richer parameter space: depending on the choice of parameters, the same analytic form can describe timelike naked singularities, regular solitons, regular black holes, Reissner-Nordström-like black holes, multi-horizon black holes, and Schwarzschild-like single-horizon black holes. 
We use this solution as a 
model to compare the geodesic structure and optical appearance of different compact object branches within a single theory.
The solution is characterized by a theory parameter, which we denote by $K$, and by the dimensionless mass parameter $\mu=M/\lambda$. 
For fixed $K$, varying $\mu$ does not describe a dynamical process, but rather a path through the parameter space of static solutions. 

We first classify the horizon structure of the solution and identify the regions of parameter space corresponding to the different branches of compact objects. 
We then analyze null and timelike geodesics, with particular emphasis on light rings, innermost stable circular orbits, and static spheres \cite{Cunha:2018acu,Gralla:2019xty,Cunha:2018gql,Cunha:2022gde,Peng:2020wun,Wei:2023bgp,Ye:2023qks,Huang:2024gtu,Huang:2025xqd,Liu:2024iec}. 
These structures determine the propagation of photons and the possible inner edge of a thin accretion disk. 
Finally, we compute the optical appearance of the compact objects by ray tracing emission from a geometrically and optically thin disk viewed face-on by a distant observer \cite{Cunningham:1973,Luminet:1979nyg,Fukue:1988,Viergutz:1993,Speith:1995,Fanton:1997,Falcke:1999pj,Gralla:2019xty,Dokuchaev:2019pcx,Gyulchev:2019tvk,Gyulchev:2020cvo,Gyulchev:2021dvt,Huang:2023yqd,Huang:2024bbs,Deliyski:2024wmt,Deliyski:2026fav,Wang:2026rae}.

A main result of our analysis is that the optical image is not determined solely by the number of horizons \cite{Gralla:2019xty,Cunha:2018gql}. 
Horizonless configurations can display shadow-like central depressions when the disk has an inner edge and the transfer function prevents photons from sampling the central region. 
Conversely, multi-horizon black holes can appear very similar to ordinary single-horizon black holes if their outer
light ring and disk structures are similar. 
The observable image is therefore mainly determined by the outer photon potential and by the location of the disk inner edge, while the inner horizon structure leaves only an indirect imprint.

To make this statement quantitative, we also introduce simple image diagnostics based on the radial intensity profile. 
We compare single-horizon and three-horizon black holes both at fixed impact parameter and after rescaling the image by the critical impact parameter. 
This allows us to distinguish true pointwise agreement of the brightness profiles from a weaker, but physically relevant, morphological degeneracy of the image structure.

The paper is organized as follows. 
We review the shift- and parity-symmetric beyond Horndeski model and the corresponding static, spherically symmetric solutions with primary scalar hair in Sec.~2. 
In Sec.~3 we specialize to the $n/s=5/2$ solution and classify its parameter space according to the horizon and regularity structure. 
In Sec.~4 we study null and timelike geodesics, focusing on light rings, innermost stable circular orbits, and static spheres. 
We compute the optical appearances of representative compact objects by ray tracing emission from a geometrically and optically thin disk in Sec.~5, and also introduce fixed-scale and shape-rescaled radial-profile diagnostics, which quantify the degree of image degeneracy between representative single-horizon and multi-horizon black holes.
We conclude in Sec.~6.

\section{Beyond Horndeski gravity}

We consider a class of scalar--tensor theories belonging to the
degenerate higher-order scalar--tensor (DHOST) framework, often referred to as beyond Horndeski theories. 
These theories extend the original Horndeski action by allowing higher-derivative interactions of the scalar field while maintaining the absence of Ostrogradsky instabilities through degeneracy conditions.

\subsection{Action and symmetries}

The action of the shift- and parity-symmetric scalar--tensor theory is defined by
\begin{equation}
S = \int d^4x \sqrt{-g} \left[
\frac{1}{2} R + \mathcal{L}_\phi
\right],
\end{equation}
where the scalar Lagrangian is given by
\begin{equation}
\mathcal{L}_\phi =
\sum_{n=0}^{\infty} c_n \, X^{n/s}
+ \beta \, X \, \Box \phi
+ \gamma \, \mathcal{L}_4^{\rm BH}.
\end{equation}
Here
\begin{equation}
X = -\frac{1}{2} \nabla_\mu \phi \nabla^\mu \phi
\end{equation}
is the kinetic term, and $\mathcal{L}_4^{\rm BH}$ denotes the beyond Horndeski interaction
\begin{equation}
\mathcal{L}_4^{\rm BH} =
\epsilon^{\mu\nu\rho}_{\ \ \ \sigma}
\epsilon^{\mu'\nu'\rho'\sigma}
\nabla_\mu \phi \nabla_{\mu'} \phi
\nabla_\nu \nabla_{\nu'} \phi
\nabla_\rho \nabla_{\rho'} \phi.
\end{equation}
The coefficients $c_n$ encode the scalar self-interactions and define the specific model considered in Ref.~\cite{Bakopoulos:2023sdm}. 
The constants $\beta$ and $\gamma$ determine the strength of the derivative couplings.

In the shift-symmetric sector, the scalar field appears only through its derivatives, and the action is invariant under
\begin{equation}
\phi \rightarrow \phi + \text{const}.
\end{equation}
As a consequence, the scalar field equation can be written in the form
of a conserved current,
\begin{equation}
\nabla_\mu J^\mu = 0,
\end{equation}
where $J^\mu$ depends on $X$ and higher derivatives of $\phi$.

Varying the action with respect to the metric yields
\begin{equation}
G_{\mu\nu} = T_{\mu\nu}^{(\phi)},
\end{equation}
where the effective energy--momentum tensor takes the schematic form
\begin{equation}
T_{\mu\nu}^{(\phi)} =
- \frac{2}{\sqrt{-g}} \frac{\delta (\sqrt{-g}\mathcal{L}_\phi)}{\delta g^{\mu\nu}}.
\end{equation}

A distinctive feature of these theories is the existence of compact objects endowed with \emph{primary scalar hair} \cite{Bakopoulos:2023sdm}. 
In contrast to secondary hair, which is determined by conserved charges, primary hair arises from nontrivial interactions in the Lagrangian and is not associated with a Gauss-law constraint.

\subsection{Static and spherically symmetric configurations}

We consider static, spherically symmetric spacetimes of the form
\begin{equation}
ds^2 = -h(r) dt^2 + \frac{dr^2}{h(r)} + r^2 d\Omega^2,
\end{equation}
together with a scalar field configuration compatible with the symmetries of the theory.

In this sector, the field equations reduce to a system of ordinary differential equations for $h(r)$ and the scalar field profile. 
Remarkably, these equations admit a family of analytic solutions characterized by a set of parameters $\{c_n\}$ entering the Lagrangian \cite{Bakopoulos:2023sdm}.

Recently, a family of static and spherically symmetric solutions with primary hair was constructed in Ref.~\cite{Bakopoulos:2023sdm}. 
These solutions depend on a set of parameters $\{c_n\}$ entering the Lagrangian and allow for a systematic expansion in powers of $(q^2/2)^{n/s}$. 
The corresponding metric function takes the general form
\begin{equation}
h(r) = 1 - \frac{2M}{r} + \left(1 + \frac{\zeta}{\gamma}
+ \frac{2\beta}{3\gamma} c_0\right)\frac{r^2}{\lambda^2}
+ \frac{2\beta}{3\gamma}\frac{r^2}{\lambda^2}
\sum_{n=1}^{\infty} c_n \left(\frac{q^2}{2}\right)^{n/s}
{}_2F_1\left(\frac{3}{2}, \frac{n}{s}; \frac{5}{2};
-\frac{r^2}{\lambda^2}\right).
\end{equation}

To obtain asymptotically flat solutions with finite ADM mass, one imposes $c_0=0$ and $\xi=-\gamma$, and further sets $c_1=0$ to avoid global monopole contributions \cite{Barriola:1989hx,Chatzifotis:2022ubq}.

For $s=2$, the lowest nontrivial contribution corresponds to $n/s=2$, yielding the black hole solution first obtained in Ref.~\cite{Bakopoulos:2023fmv}. 
This solution describes asymptotically flat black holes with primary scalar hair and exhibits deviations from the Schwarzschild geometry at intermediate scales.

The properties of this solution have been further investigated in Ref.~\cite{Erices:2024lci}, where its quasinormal modes, thermodynamic behavior, and optical images were analyzed. 
These studies show that the presence of primary scalar hair can lead to observable deviations in the dynamical and optical properties of the black hole spacetime.
Here we consider the next contribution in the expansion, corresponding to $n/s = 5/2$.
Unlike the $n/s=2$ case, this solution exhibits a much richer structure.

\section{Phases of the solution}

Following Ref.~\cite{Bakopoulos:2023sdm}, we consider the next contribution in the expansion, corresponding to $n/s = 5/2$. 
The resulting metric function is given by
\begin{equation}
\label{ds}
h(r) = 1 - \frac{2M}{r} + \frac{\sqrt{2}\beta c_{5/2} q^5}{3\gamma\,\lambda\, r}
- \frac{\sqrt{2}\beta c_{5/2} q^5}{3\gamma}
\frac{r^2/\lambda^2}{(1 + r^2/\lambda^2)^{3/2}}.
\end{equation}
Depending on the values of the parameters, it can describe a wide variety of compact objects, including naked singularities, regular black holes, Schwarzschild-like or RN-like black holes, multi-horizon black holes, and solitons.

This extended solution space provides an ideal framework to investigate how the geometric and optical properties of compact objects change as the mass parameter is varied within a fixed theory. 
In particular, it allows us to explore transitions between qualitatively distinct configurations, which is the main focus of this work.

\subsection{General properties}

The solution \eqref{ds} is asymptotically flat with ADM mass $M$. 
The scalar curvature $R$ and the Kretschmann scalar $\cal K = R_{\mu\nu\rho\sigma}R^{\mu\nu\rho\sigma}$ are given by
\bea
R &=&-h''(r)-\frac{4h'(r)}{r}-\frac{2(h(r)-1)}{r^2},\nn\\
{\cal K} &=&h''(r)^2+\frac{4h'(r)}{r^2}+\frac{4(h(r)-1)^2}{r^4}.
\eea
One can verify that the scalar curvature $R$ remains finite everywhere. 
However, for $M/\lambda \neq \beta c_{5/2} q^5 / (3\sqrt{2}\gamma)$, the Kretschmann scalar $\cal K$ diverges at $r=0$, indicating a curvature singularity. 
In contrast, when $M/\lambda = \beta c_{5/2} q^5 / (3\sqrt{2}\gamma)$, $\cal K$ becomes finite at $r=0$, implying that the spacetime is regular. 
Following Ref.~\cite{Bakopoulos:2023sdm}, we set $\beta = \gamma = -1$ without loss of generality. 
Introducing dimensionless variables
\be
x=\frac{r}{\lambda},\qquad
\mu=\frac{M}{\lambda},\qquad
\kappa=\frac{\sqrt{2}}{3}\,\frac{c_{5/2}q^5}{\lambda},
\ee
the metric function becomes
\be
h(x)=1-\frac{2\mu-\kappa}{x}-\kappa\frac{x^2}{(1+x^2)^{3/2}}.
\label{dsnew}
\ee
The regular branch is obtained when the $1/r$ singular term vanishes, namely
\be
\mu=\frac{\kappa}{2}.
\ee
Equivalently, defining 
\be
K:=\frac{c_{5/2}q^5}{\lambda},
\ee
the regular line is
\be
\mu=\frac{K}{3\sqrt2}.
\ee

The metric function $h(x)$ in \eqref{dsnew} depends on two parameters, $\kappa$ and $\mu$. 
The mass parameter $\mu$ is an integration constant of the solution, while the parameter $\kappa$ enters the Lagrangian and thus represents a theory parameter. 
Fixing $\kappa$ (or equivalently $K$) specifies the underlying theory, while varying $\mu$ generates different spacetime configurations: 
naked singularities (no horizon), Schwarzschild-like black holes (one horizon), regular black holes (one degenerate horizon or two horizons),  RN-like black holes (two horizons), multi-horizon black holes (three horizons), and solitons (no horizon). 

To classify these configurations, we examine when the horizon structure changes.
This occurs when $h(x)$ develops a double root:
\be
h(x)=0,\qquad h'(x)=0.
\ee
Solving these equations parametrically yields
\be
\kappa(x)=\frac{(1+x^2)^{5/2}}{3x^2},
\qquad
\mu(x)=\frac12\left[\kappa(x)+\frac{x(2-x^2)}{3}\right].
\ee
In terms of $K=\dfrac{3}{\sqrt2}\kappa$, this becomes
\be
K(x)=\frac{(1+x^2)^{5/2}}{\sqrt2\,x^2},
\qquad
\mu(x)=\frac12\left[\frac{\sqrt2}{3}K(x)+\frac{x(2-x^2)}{3}\right].
\ee
This parametric curve defines the extremal configurations and consists of two branches for $K>K_1$, which we denote by $\mu_-(K)$ and $\mu_+(K)$.

The two characteristic values are
\be
K_1=\frac{25\sqrt{30}}{36}\approx 3.80363,
\qquad
K_2=\frac{9\sqrt6}{4}\approx 5.51135.
\ee
Here, $K_1$ is the minimum value of the extremal curve, where the two branches emerge, while $K_2$ is the value where the regular line $\mu=K/(3\sqrt2)$ intersects the upper extremal curve and the regular solution becomes extremal.

\subsection{Classification of the mass paths at fixed $K=c_{5/2}q^5/\lambda$}

As the mass parameter $\mu=M/\lambda$ varies at fixed $K$, three qualitatively distinct mass paths arise. 

\begin{figure}[H]
\centering{\includegraphics[width=0.5\textwidth]{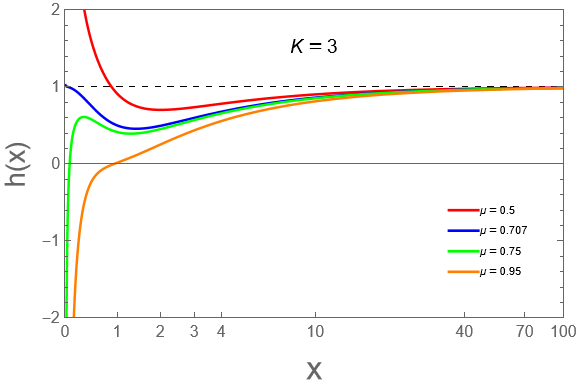}}
\caption{\it The evolution of path A.}\label{pathc}
\end{figure}

\paragraph{Path A: $0<K<K_1$.}
In this case, there is only one transition at the regular point. 
As the mass increases, the sequence passes from a naked singularity to a regular soliton, which then becomes a single-horizon black hole.
The resulting black hole is not necessarily Schwarzschild-like, since the metric function $h(x)$ need not be monotonically increasing outside the horizon. 
This non-monotonic behavior can lead to distinct physical properties. 
The corresponding evolution is illustrated in Fig.~\ref{pathc}.

\begin{figure}[H]
\centering{\includegraphics[width=0.5\textwidth]{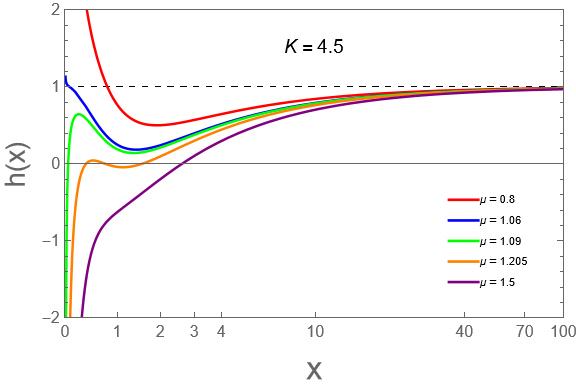}}
\caption{\it The evolution of path B.}\label{pathb}
\end{figure}

\paragraph{Path B: $K_1<K<K_2$.}
In this regime, a three-horizon window appears, while the regular solution remains a soliton.
As an example, for $K=4.5$, the sequence is as follows.
A low-mass naked singularity first becomes a star-like soliton at $\mu=1.060$. 
A small increase in mass gives rise to a single-horizon black hole. 
As the mass increases further a three-horizon black hole appears in the interval $1.180<\mu<1.217$. 
Finally, for $\mu>1.217$ the solution becomes a Schwarzschild-like black hole with a single horizon. 
We show this type of evolution in Fig.~\ref{pathb}.

\begin{figure}[H]
\centering{\includegraphics[width=0.5\textwidth]{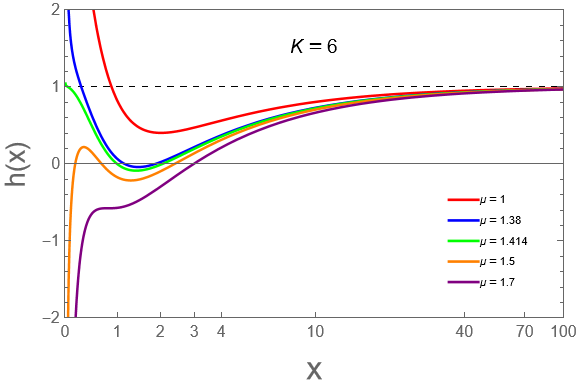}}
\caption{\it The evolution of path C.}\label{patha}
\end{figure}

\paragraph{Path C: $K>K_2$.}
For $K>K_2$, the regular point lies inside the black-hole region, and therefore no regular soliton exists.  
For example, when $K=6$ the solution starts as a timelike naked singularity at small $\mu$.
As the mass increases, two horizons form, corresponding to a RN-like black hole.  
At $\mu=\ft{\sqrt{2}}{6} K$, the singularity disappears and the solution becomes a regular black hole. 
For larger masses a three-horizon black hole appears in the interval $\mu\in (1.415,1.542)$, before transitioning to a single-horizon black hole for $\mu>1.542$. 
This evolution is illustrated in Fig.~\ref{patha}.

\begin{figure}[H]
\centering{\includegraphics[width=0.5\textwidth]{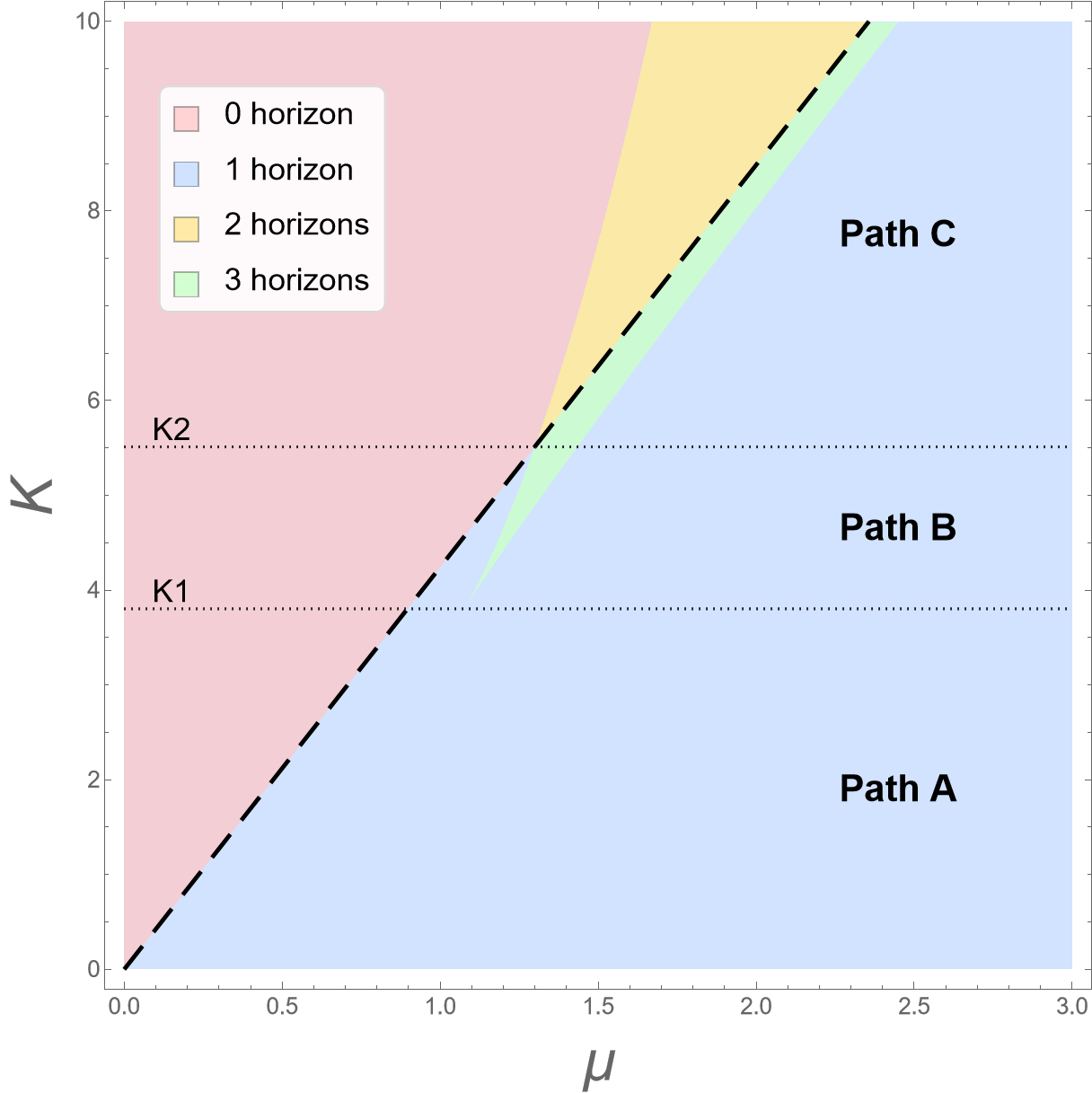}}
\caption{\it The parameter space of the solution.}\label{para}
\end{figure}

In Fig.~\ref{para}, we show the parameter space of the solution, which illustrates the three possible mass paths with the theory parameter $K$ fixed.

\section{Geodesics and the effective potential}

The horizon and regularity classification discussed above provides only a first characterization of the different compact-object branches. 
To understand their optical appearance, one also has to analyze the geodesic structure outside the compact object. 
In particular, unstable light rings control the strong deflection of photons, while timelike circular orbits determine possible inner scales of a thin accretion disk. 
We therefore study null and timelike geodesics in the dimensionless metric introduced above, focusing on light rings, ISCOs, and static spheres \cite{Cardoso:2014sna}.

The geodesics are described by the world line\footnote{In this section we use the dimensionless radial coordinate $x=r/\lambda$, and all lengths such as the impact parameter $b$ are measured in units of the length scale $\lambda$.}
\be
x^{\mu}(\sigma)=\{t(\sigma),x(\sigma), \theta(\sigma), \varphi(\sigma)\},
\ee
where $\sigma$ denotes the affine parameter. 
The tangent vector of the world line is given by 
\be
\xi^\mu=\frac{d x^\mu(\sigma)}{d\sigma},
\ee
which satisfies
\be\label{tan}
\xi^{\mu}\xi_\mu=\epsilon,
\ee
where $\epsilon=0$ for massless particles and $\epsilon=-1$ for massive particles.
This leads to 
\be\label{rays}
-h(x)(\frac{dt}{d\sigma})^2+h(x)^{-1}(\frac{dx}{d\sigma})^2+x^2 (\frac{d\varphi}{d\sigma})^2=\epsilon,
\ee
where we have chosen the equatorial plane $\theta=\ft{\pi}{2}$ without loss of generality.  
There are two conserved quantities,
\be
E=h(x)\frac{dt}{d\sigma}, \qquad L=x^2\frac{d\varphi}{d\sigma}.
\ee
Then Eq.~\ref{rays} reduces to 
\be\label{neo}
\frac{dx}{d\varphi} = \pm \, x^2 \sqrt{\frac{1}{b^2} - V_{geo,\epsilon}}.
\ee
where the impact parameter $b$ and the effective potential are given by
\be
b=\frac{L}{E}, \qquad V_{geo,\epsilon} = \frac{h(x)}{x^2} - \frac{\epsilon h(x)}{L^2}.
\ee

\subsection{Massless particles and light rings}

\begin{figure}[H]
\centering{\includegraphics[width=0.5\textwidth]{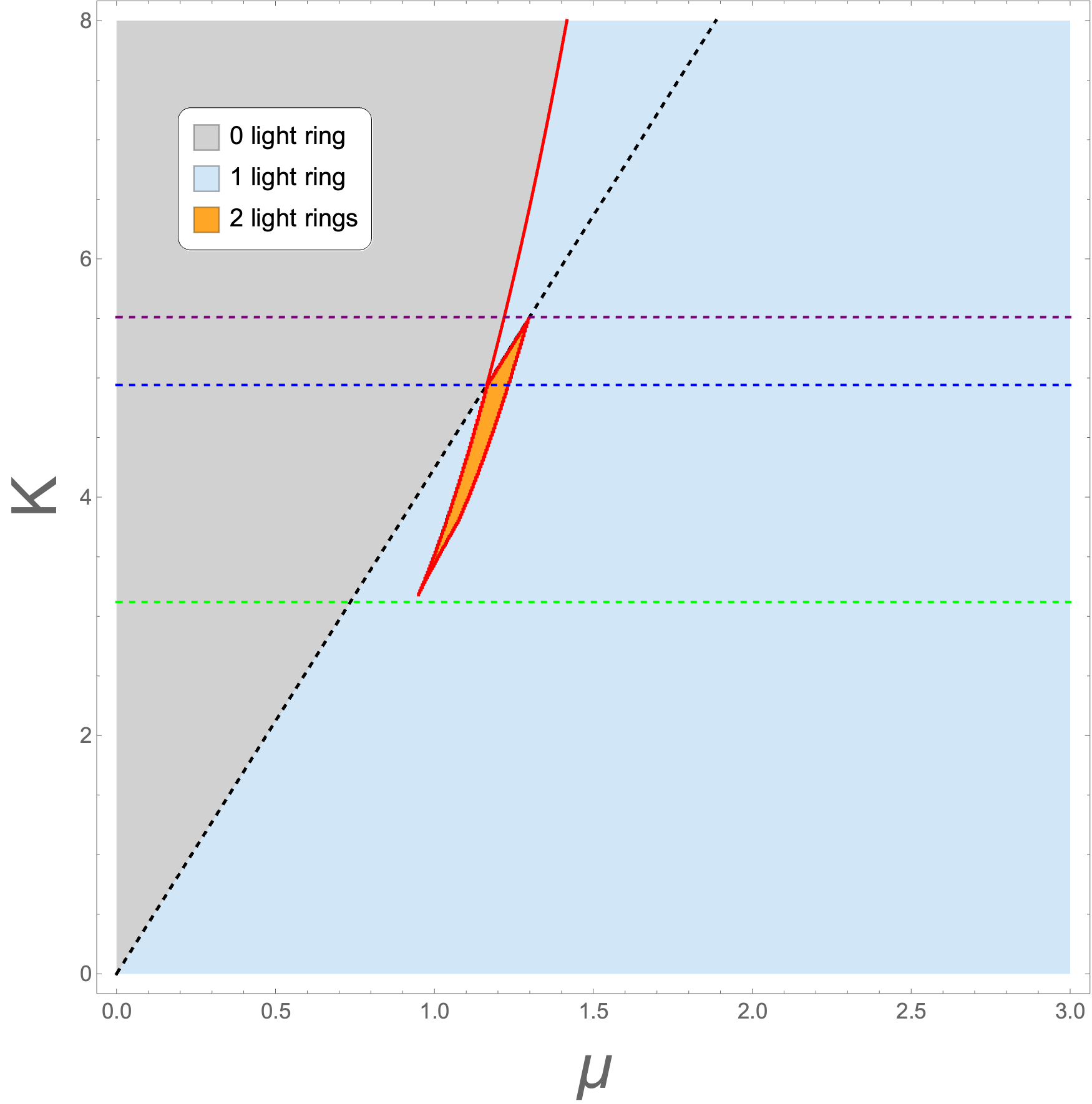}}
\caption{\it The parameter space of light rings. The green dashed line represents $K = 3.087$, the blue dashed line represents $K = 4.942$, and the purple dashed line represents $K = 5.511$. }
\label{light}
\end{figure}

For massless particles ($\epsilon=0$), the effective potential becomes
\be
V_{geo,0}=\frac{h(x)}{x^2}.
\ee
A light ring corresponds to an unstable orbit of massless particles and satisfies 
\be
\frac{\partial V_{geo,0}}{\partial x}|_{x=x_{LR}} = 0, \qquad
\frac{\partial^2 V_{geo,0}}{\partial x^2}|_{x=x_{LR}} < 0.
\ee
We emphasize that $N_{\rm LR}$ denotes the number of unstable outer
light rings, i.e. local maxima of $V_{\rm geo,0}$. Stable light rings, corresponding to local minima of the same potential, are not included in this count. Therefore the changes in $N_{\rm LR}$ in Fig.~\ref{light} should not be
interpreted as contradicting the usual topological light-ring counting
arguments, which apply to the full set of light rings under additional
regularity and energy-condition assumptions \cite{Cunha:2022gde}. In most physically relevant cases, black holes possess at least one unstable light ring.
However, the existence and number of light rings depend sensitively on the structure of the metric function $h(x)$ \cite{Cunha:2017qtt,Hod:2017zpi,Cunha:2020azh}.

For evolution along path A (i.e., $0< K < K_1$), no light rings exist during the naked singularity phase. 
As the mass increases and the object becomes a single-horizon black hole, generally, at least one light ring appears. 
However, in certain regions of the $(K, \mu)$ parameter space, these single-horizon black holes can admit two light rings.

For evolution along path B (i.e., $K_1< K < K_2$), the solution may possess a light ring even during the naked singularity phase, and one or two light rings in the single-horizon black hole regime.  
In the three-horizon black hole regime, we find that only a single light ring is present.
Consequently, the number of light rings does not vary monotonically with the mass of the solution.
This behavior reflects the appearance and disappearance of extrema in the effective potential $V_{geo,0}$ as the spacetime geometry changes.

\begin{figure}[H]
\centering
\includegraphics[width=0.4\textwidth]{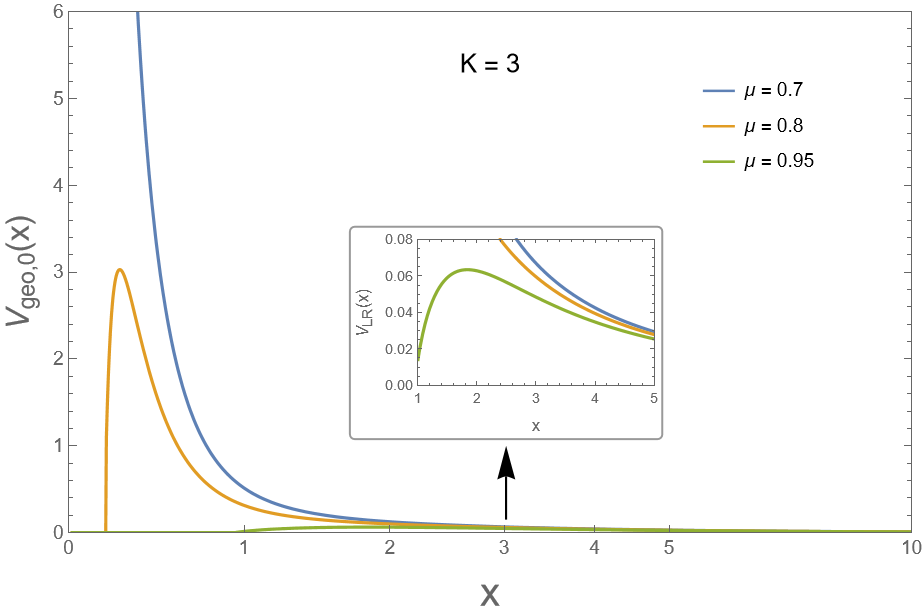}\qquad
\includegraphics[width=0.4\textwidth]{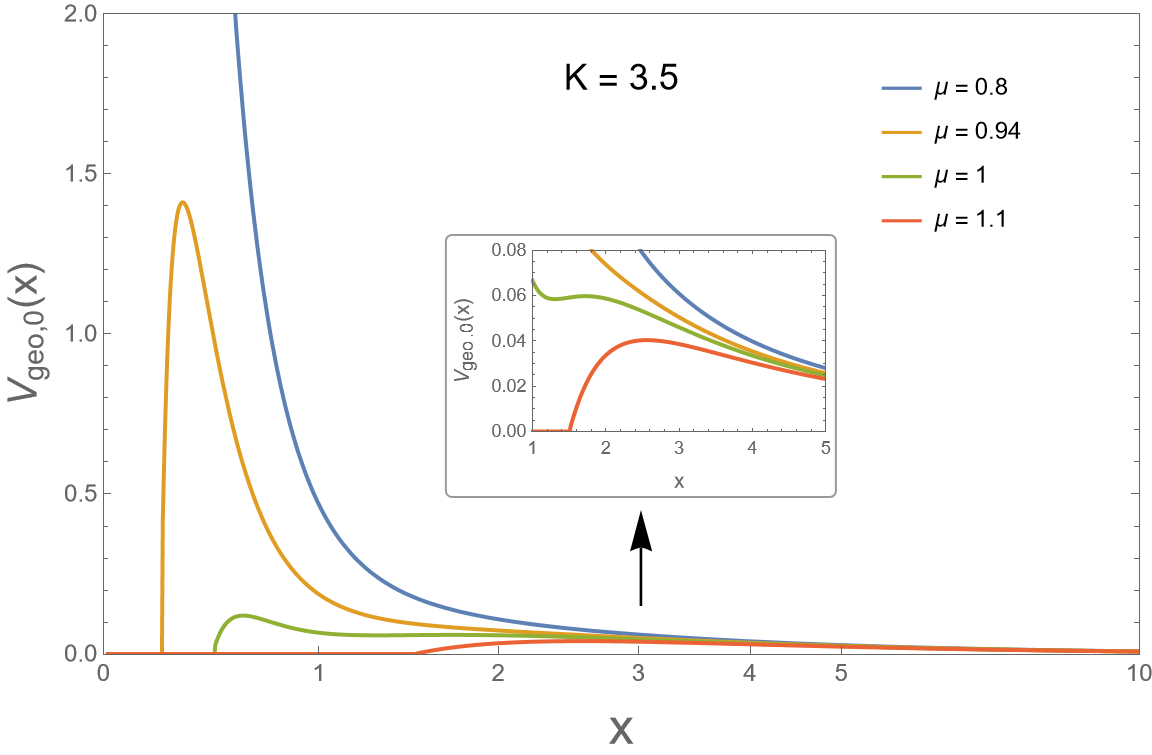}\qquad
\includegraphics[width=0.4\textwidth]{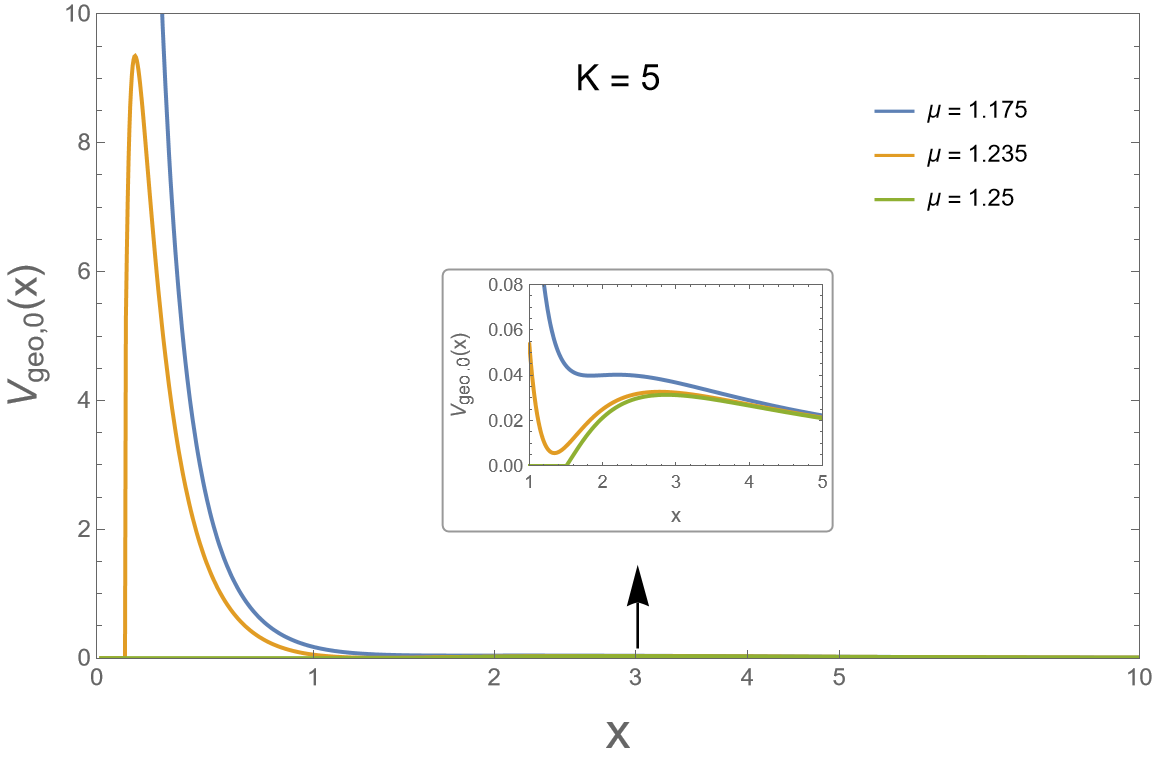}\qquad
\includegraphics[width=0.4\textwidth]{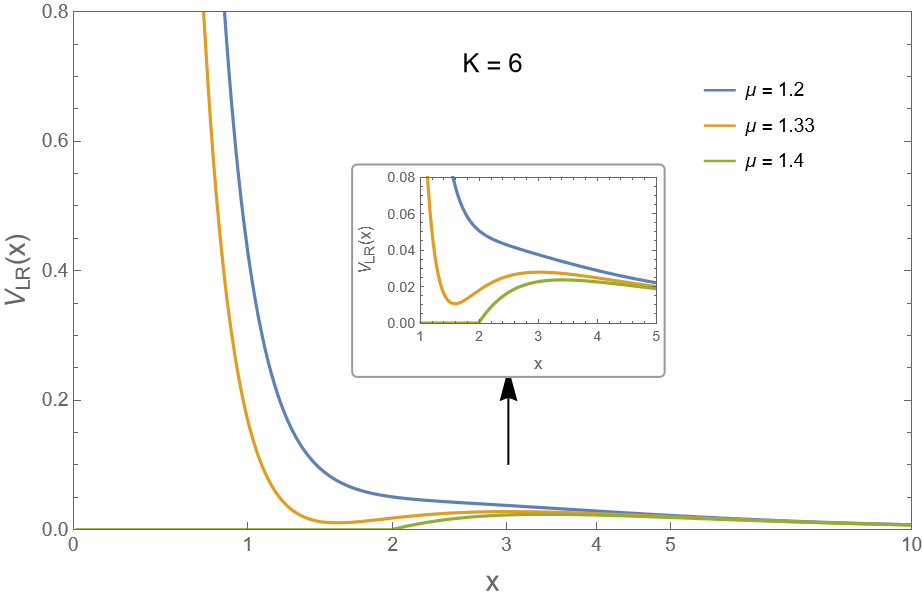}
\caption{\it Effective potentials for four different cases: $K=3$ (for path A), $K=3.5$ and $K=5$ (for path B), and $K=6$ (for path C). 
Insets highlight the fine structure near the potential maximum.}
\label{VK3.5}
\end{figure}

For evolution along path C (i.e., $K>K_2$), the solution always exhibits at least one light ring whether in the naked singularity phase or in the multi-horizon black hole phases.

\begin{figure}[H]
\centering
\includegraphics[width=0.4\textwidth]{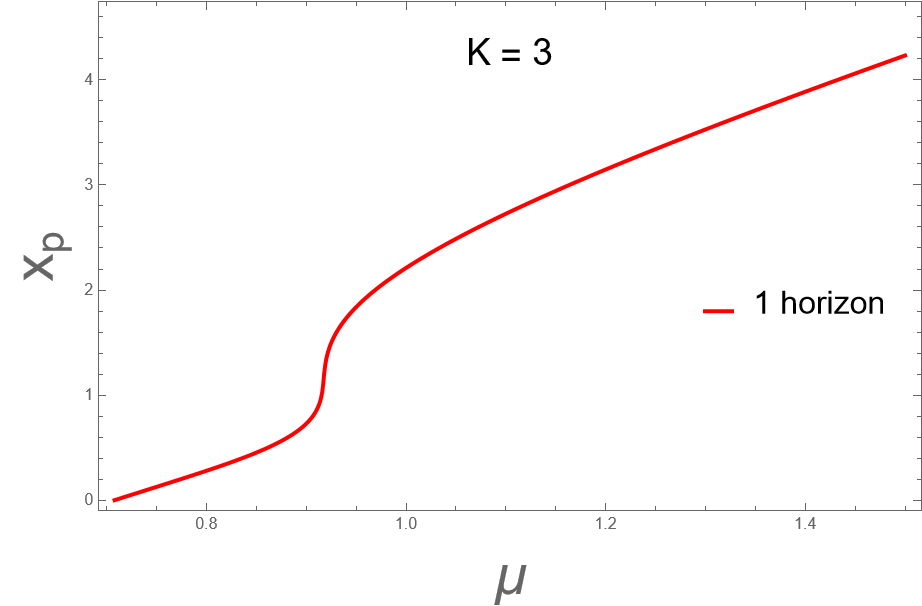}\qquad
\includegraphics[width=0.4\textwidth]{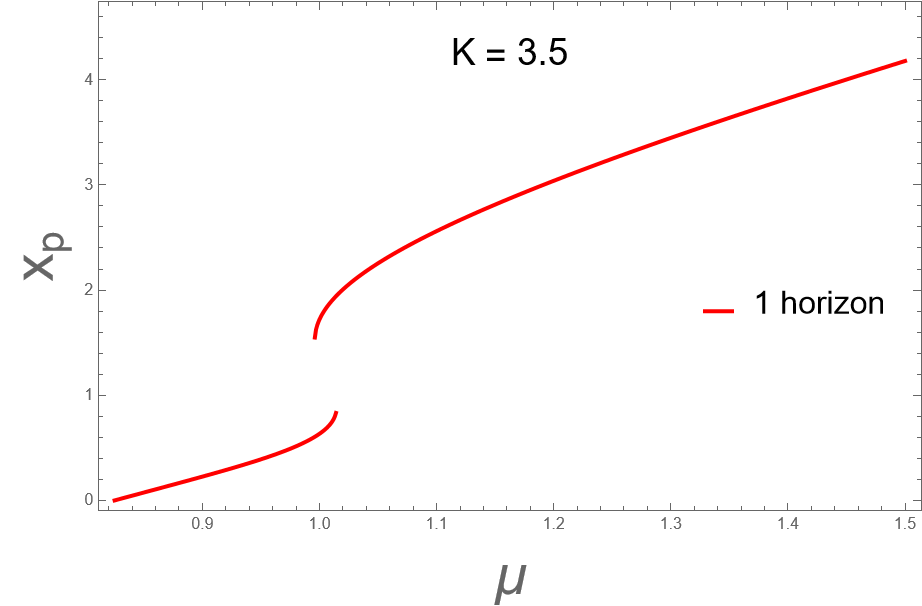}\qquad
\includegraphics[width=0.4\textwidth]{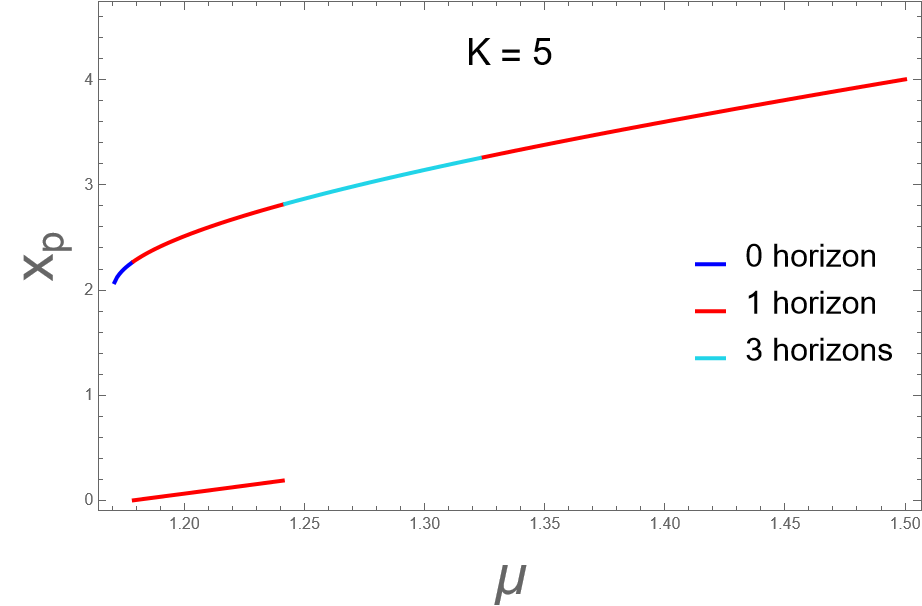}\qquad
\includegraphics[width=0.4\textwidth]{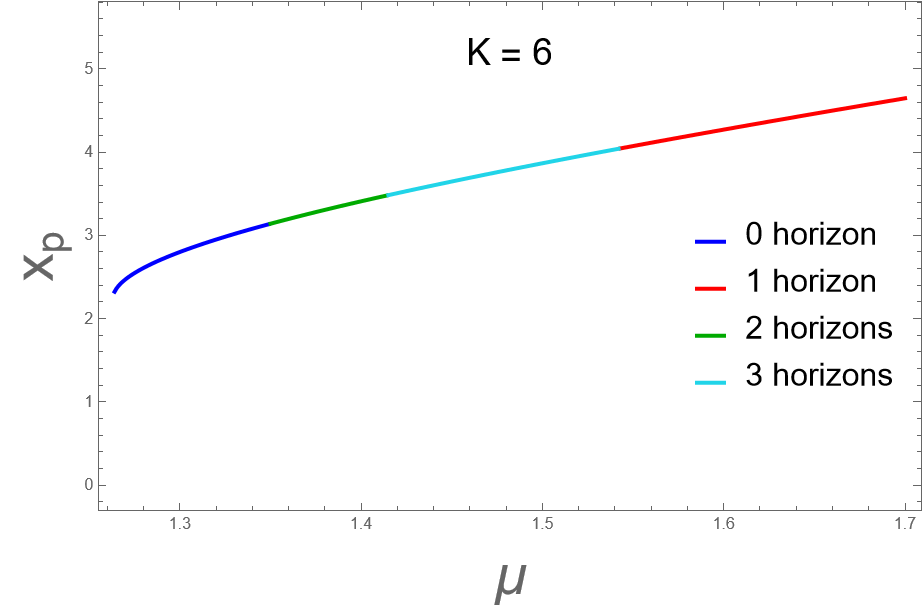}
\caption{\it Light ring radius $x_p$ as a function of the parameter $\mu$ for four representative cases with $K = 3, 3.5, 5, 6$. 
The curves illustrate how the number and location of light rings change with solution mass $\mu$, corresponding to spacetimes with 0, 1, 2, or 3 horizons as labeled.}
\label{rpK3.5}
\end{figure}

The distribution of light ring numbers in the $(K, \mu)$ parameter space, is shown in Fig.~\ref{light}. 
Comparing with Fig.~\ref{para}, we observe that the regions with multiple light rings partially overlap with, but do not coincide with, the multi-horizon region.
To illustrate the evolution more explicitly, we select representative $K$ values and show the effective potential $V_{geo,0}$ for different masses along the evolution paths in Fig.~\ref{VK3.5}. 
The corresponding variation of the light ring radius with $\mu$ is presented in Fig.~\ref{rpK3.5}.

\subsection{Massive particles, ISCO and static sphere}

For massive particles, we have $\epsilon=-1$ and the effective potential becomes
\be\label{efftime}
V_{geo,-1}=\frac{h(x)}{x^2}+\frac{h(x)}{L^2}.
\ee
For different values of the angular momentum $L$, the shape of the effective potential varies. 
For certain non-vanishing values of $L$, stable circular orbits exist, with the innermost such orbit referred to as the innermost stable circular orbit (ISCO). 
In many studies of optical images of black holes, the accretion disk is assumed to extend from the ISCO. 
The ISCO is determined by 
\be
\frac{\partial V_{geo,-1}}{\partial x}|_{x=x_{ISCO}}=0, \qquad \frac{\partial^2 V_{geo,-1}}{\partial x^2}|_{x=x_{ISCO}}=0,
\ee
or equivalently by \cite{Gao:2023mjb,Zhang:2024hix}
\be
h(x_{ISCO})h''(x_{ISCO})-2h'(x_{ISCO})^2+\frac{3}{x_{ISCO}}h(x_{ISCO})h'(x_{ISCO})=0.
\ee

\begin{figure}[H]
\centering{\includegraphics[width=0.5\textwidth]{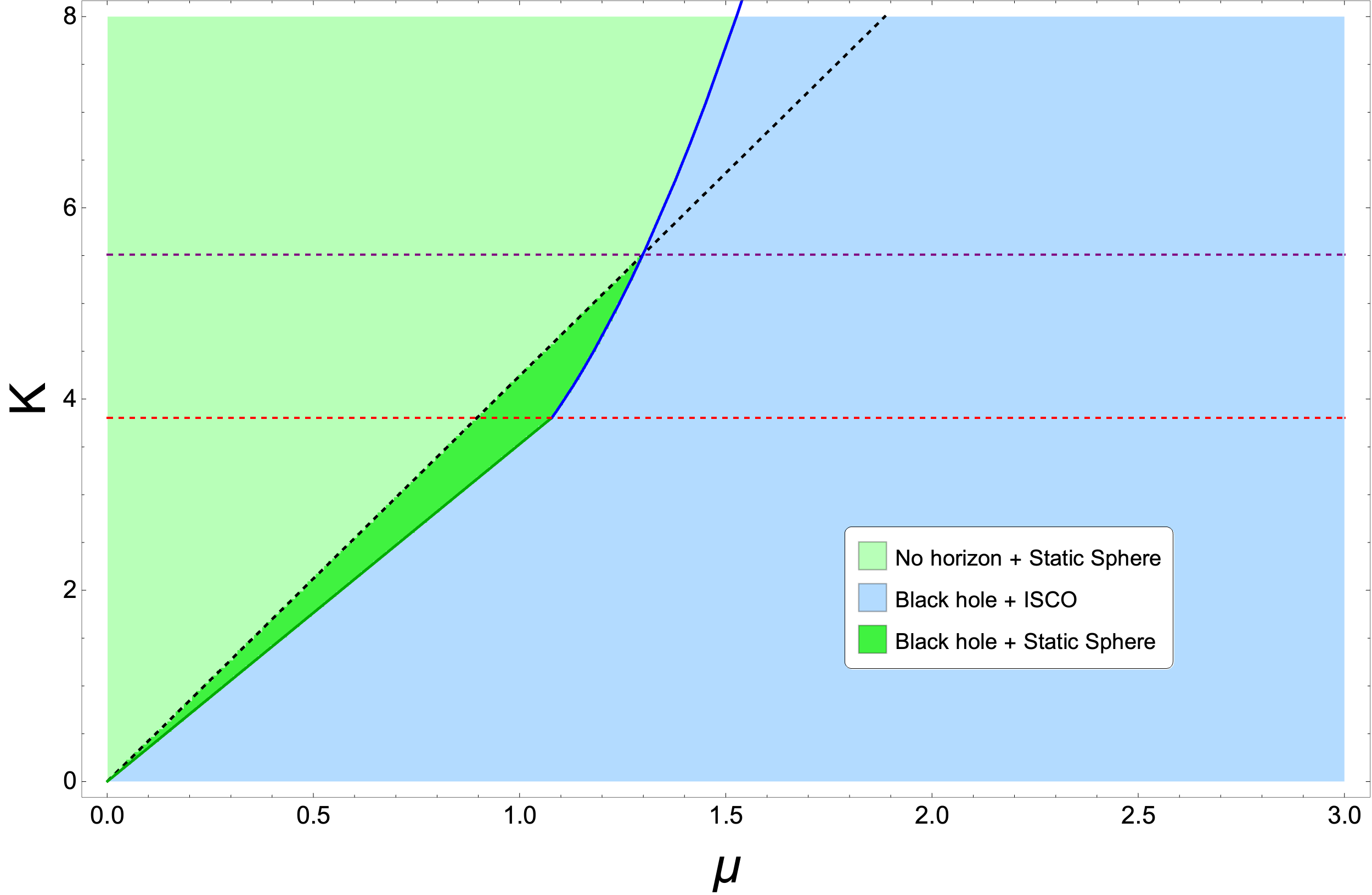}}
\caption{\it The parameter space of ISCO and static sphere. The red dashed line represents $K = 3.803$, and the purple dashed line represents $K = 5.511$. }\label{isco}
\end{figure}

However, some compact objects, such as naked singularities, may not have an ISCO. 
They may possess a so-called static sphere, which is also the innermost orbit but with $L=0$ \cite{Huang:2024bbs,Wei:2023bgp}. 
In many examples, black-hole disks are modeled with an ISCO inner edge, whereas horizonless or naked-singularity spacetimes may instead admit a static sphere that can serve as an inner disk scale.
Interestingly, we find that in a narrow region of the $(\mu,K)$ parameter space, black holes possess a static sphere rather than an ISCO.
We present the corresponding picture in Fig.~\ref{isco}.

Now we explain how to determine the boundaries. 
Firstly, an exterior static sphere requires
\be
h'(x_s)=0,
 \qquad h(x_s)>0.
\ee
The first condition gives rise to 
\be\label{cri}
\frac{\mu}{K}=\frac{\sqrt{2}}{6}
\bigg(1+\frac{x_s^3(2-x_s^2)}{(1+x_s^2)^{5/2}}\bigg).
\ee
The stable and unstable static spheres are defined by $h''(x_s)>0$ and $h''(x_s)<0$, respectively. 
The boundary condition $h''(x_s)=0$ gives rise to\footnote{This relates to the starting point of the blue curve in Fig.~\ref{isco}.}
\be
x_s=\sqrt{\frac{2}{3}}.
\ee
Substituting $x_s$ into \eqref{cri}, we get
\be
\left.\frac{\mu}{K}\right|_{critical}=\frac{\sqrt{2}(125+8\sqrt{10})}{750}\simeq 0.283405.
\ee
Thus the stable static-sphere branch exists only for
\be
0<\frac{\mu}{K}<0.283405.
\ee

Furthermore, the boundary at which the static sphere lies exactly on a horizon is determined by
\be
 h(x_s)=0,
 \qquad h'(x_s)=0.
 \ee
The condition $h(x_s)=0$ gives
\be
K=\frac{(1+x_s^2)^{5/2}}{\sqrt{2}\,x_s^2}.
\ee
The second condition $h'(x_s)=0$ is equivalent to \eqref{cri}. 
Hence the blue boundary curve in the $(\mu,K)$ plane is fixed.

\section{Optical images with a thin accretion disk}

\begin{figure}[H]
\centering{\includegraphics[width=0.82\textwidth]{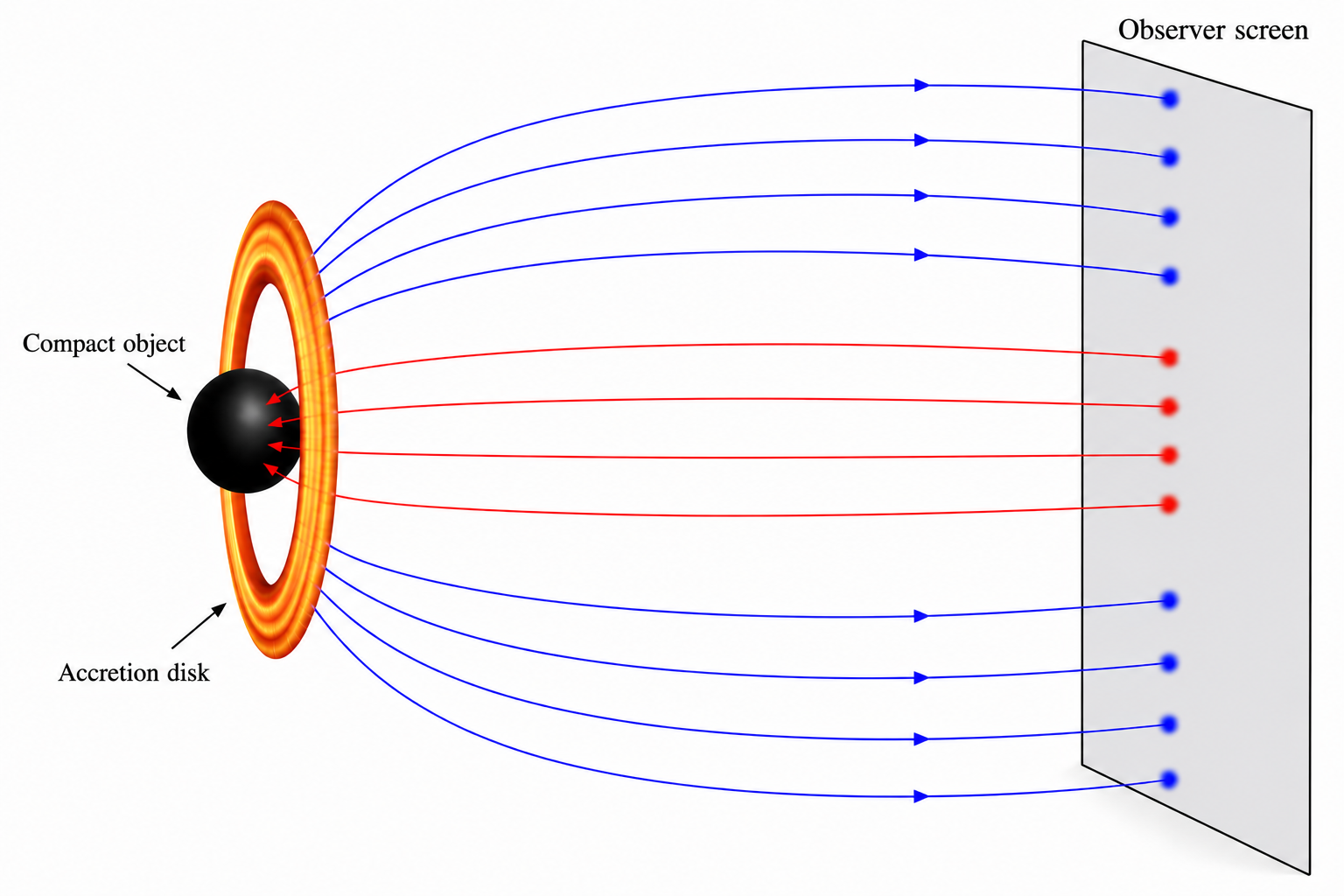}}
\caption{\it Schematic picture of the ray-tracing setup. 
The compact object is surrounded by a thin accretion disk. 
Light rays emitted from the disk are deflected by the compact object and then reach the observer's screen at infinity.}
\label{setup}
\end{figure}

We consider optical images of compact objects illuminated by an optically and geometrically thin disk. 
Although our disk model is deliberately simple, it captures the main lensing features relevant for comparing different compact-object branches and for identifying possible degeneracies in horizon-scale images.
The accretion disk, extending outward from the ISCO or the static sphere, serves as the light source, and we compute the image seen by a distant observer at infinity. 
We focus on a simple configuration, where the accretion disk is parallel to the observer's screen, as commonly adopted in the literature (see e.g.~\cite{Gralla:2019xty,Huang:2024bbs,Bambi:2012tg,Zeng:2020dco,Zhang:2023okw,Lim:2025cne}).
The observer is placed at spatial infinity, as illustrated in Fig.~\ref{setup}.

We use the same ray-tracing method as in Refs.~\cite{Gralla:2019xty,Bambi:2012tg,Li:2024kyv}. 
Since the spacetime and the disk configuration are both spherically symmetric and face-on, the final image is circularly symmetric. 
Therefore, once the observed intensity $I_\text{obs}(b)$ is known, the two-dimensional image is obtained by rotating the one-dimensional intensity profile around the center of the screen.

Following Ref.~\cite{Gralla:2019xty}, for a light ray with impact parameter $b$, the observed intensity $I_{obs}$ is obtained from the emitted intensity $I_\text{em}$ as
\be
I_\text{obs}(b)=\sum_n h^2(x_n(b))I_\text{em}(x_n(b)),
\ee
where $x_n(b)$ is the transfer function, denoting the $n$-th intersection
between the light ray and the accretion disk. 
The factor $h^2(x_n)$ is the redshift factor for the static metric considered here. 
If the light ray is captured by a horizon, or if it turns around before intersecting the disk, the corresponding term is absent. 
Thus the zeros of $I_\text{obs}$ are as important as its peaks, since they determine the dark regions and their boundaries in the image.

The inner edge of the accretion disk is chosen according to the structure of timelike geodesics. 
For the black holes with an ISCO, the disk extends from $x_{ISCO}$ to infinity. 
For the other compact objects without ISCOs but with a static sphere, the disk extends from $x_s$ to infinity. 
In the numerical images below we use a simple emission profile which decreases from the inner edge \cite{Li:2021riw,Guerrero:2021ues,Bambhaniya:2021ugr,Rosa:2022tfv,Guerrero:2022qkh,Guerrero:2022msp}, 
\be
I_\text{em}(x)=I_0\left(\frac{x_{in}}{x}\right)^4
\frac{1+\tanh[50(x-x_{in})]}{2},
\ee
where $x_{in}=x_{ISCO}$ or $x_s$, depending on the case considered. 
The overall normalization $I_0$ is irrelevant for the qualitative appearance and is chosen only for convenience.

We also show the total deflection angle $\phi_\text{tot}$ as a function of
the impact parameter. 
For a scattering light ray with turning point $x_0$,
\be
\phi_\text{tot}(b)=2\int\limits_{x_0}^{\infty}
\frac{d x}{x^2\sqrt{\ft{1}{b^2}-\frac{h(x)}{x^2}}}.
\ee
When $b$ approaches the critical impact parameter of an unstable light ring, $\phi_\text{tot}$ diverges. 
This divergence indicates that the corresponding light rays wind many times around the compact object and give rise to a sequence of bright rings in the image. 
In the figures below the horizontal red line denotes $\phi_\text{tot}=\pi$. 
For black holes, the curve is plotted only for rays returning to infinity.

The deflection angle provides a useful diagnostic of the image. 
In the face-on geometry, the first intersection with the disk gives the direct image of the disk. 
Light rays that have accumulated a larger azimuthal angle can intersect the disk again, producing a demagnified image of the opposite side of the disk. 
These secondary contributions form the lensing ring. 
If the light ray winds very close to an unstable light ring, further intersections occur and produce very narrow higher-order ring contributions. 
In practice, the higher-order rings are difficult to distinguish in the two-dimensional image, because the corresponding intervals of $b$ become very narrow.
Nevertheless, they are visible as sharp peaks in $I_\text{obs}(b)$, and they are correlated with the steep growth or divergence of $\phi_\text{tot}(b)$.

It is important to distinguish the dark central region in the image from the ideal critical curve of a black hole. 
For a thin disk, the boundary of the dark region is also controlled by the inner edge of the disk and by the transfer function. 
Hence, a horizonless object may show a shadow-like dark region if the disk has an inner boundary and the light rays from the central region do not intersect the disk. 
Conversely, a black hole image can have a dark region whose radius is larger than the critical impact parameter, because the disk emission starts from the ISCO or from the static sphere. 

\subsection{Optical images along path A}

\begin{figure}[H]
\centering
\includegraphics[width=0.30\textwidth]{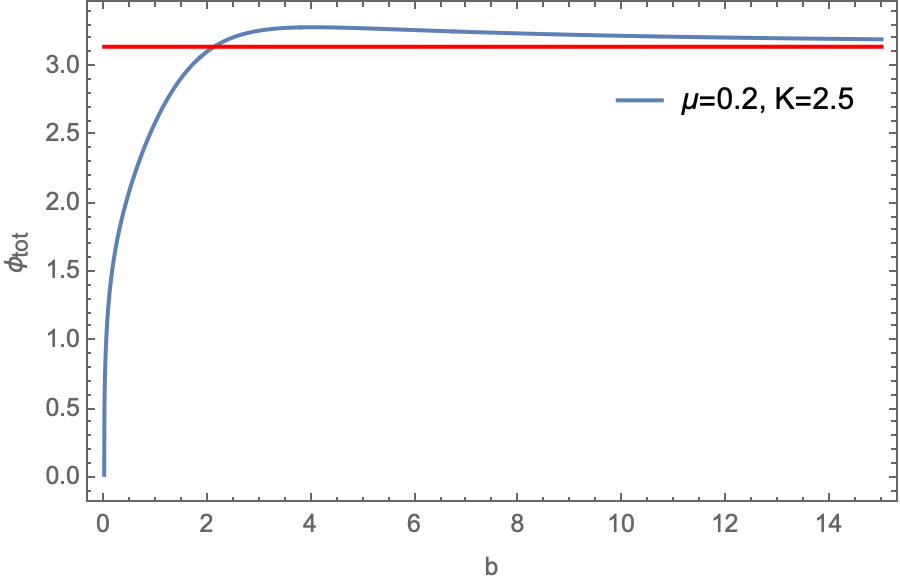}\hspace{0.025\textwidth}%
\includegraphics[width=0.30\textwidth]{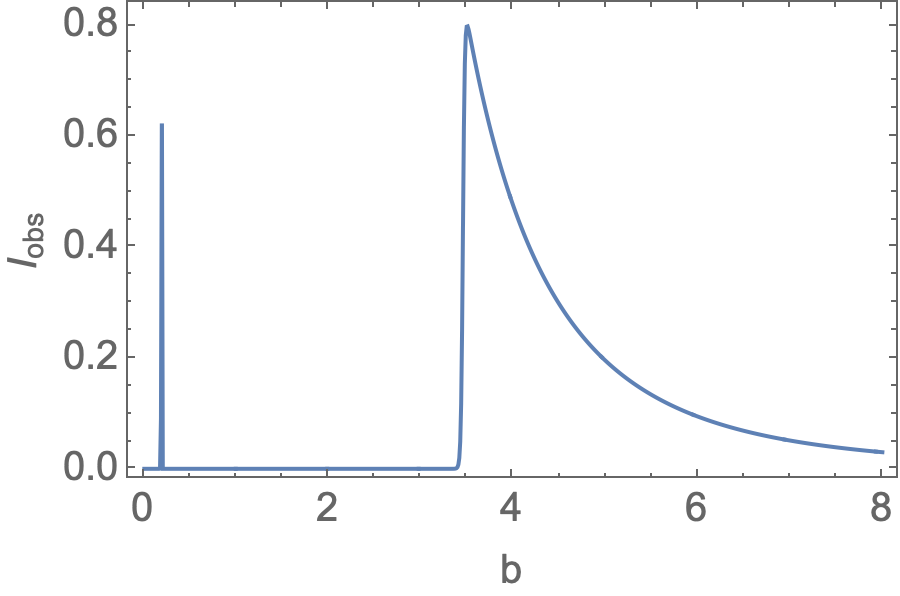}\hspace{0.025\textwidth}%
\includegraphics[width=0.30\textwidth]{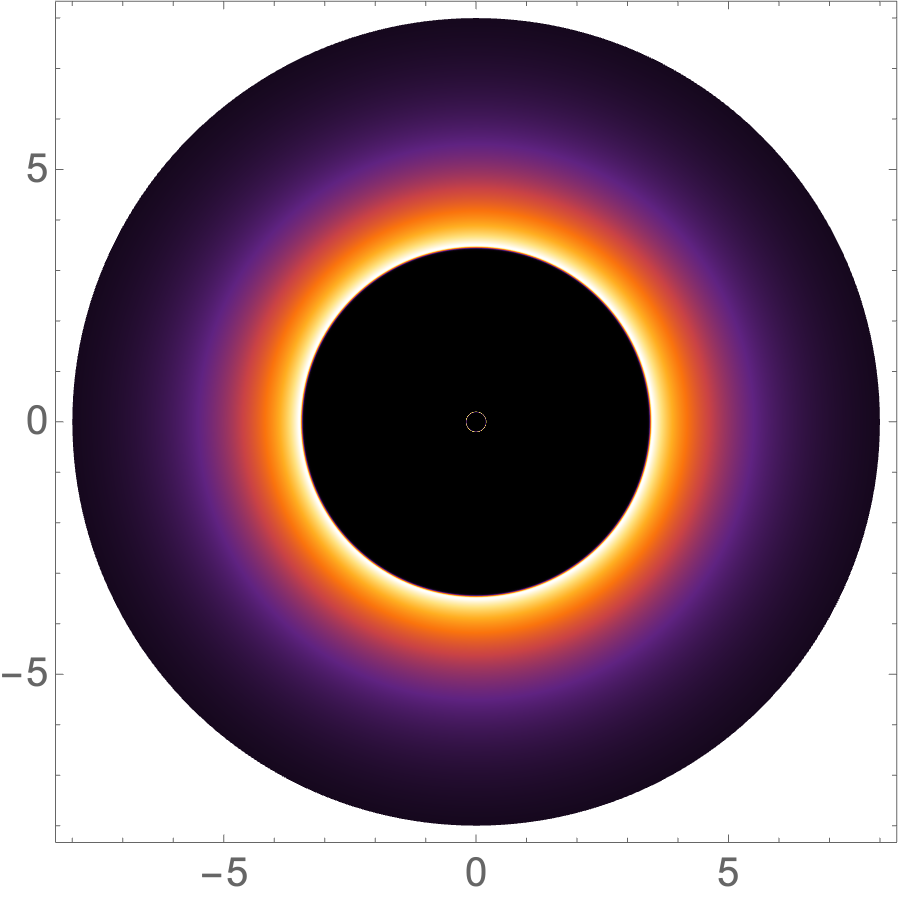}\par\vspace{1mm}
\includegraphics[width=0.30\textwidth]{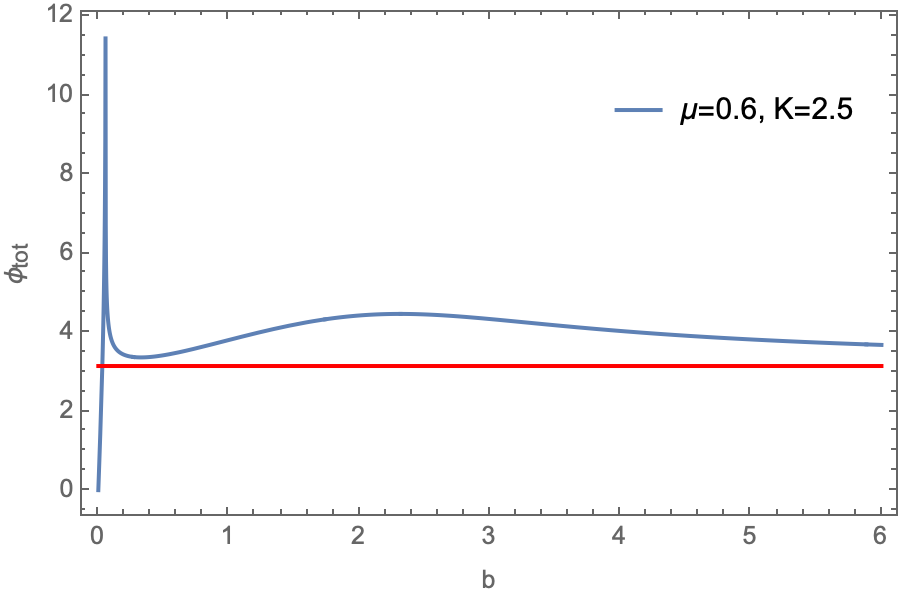}\hspace{0.025\textwidth}%
\includegraphics[width=0.30\textwidth]{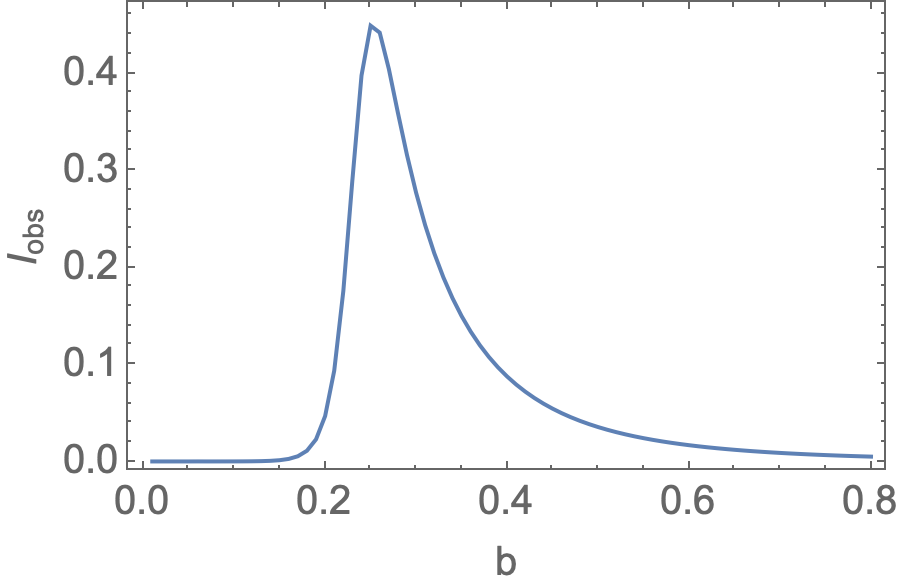}\hspace{0.025\textwidth}%
\includegraphics[width=0.30\textwidth]{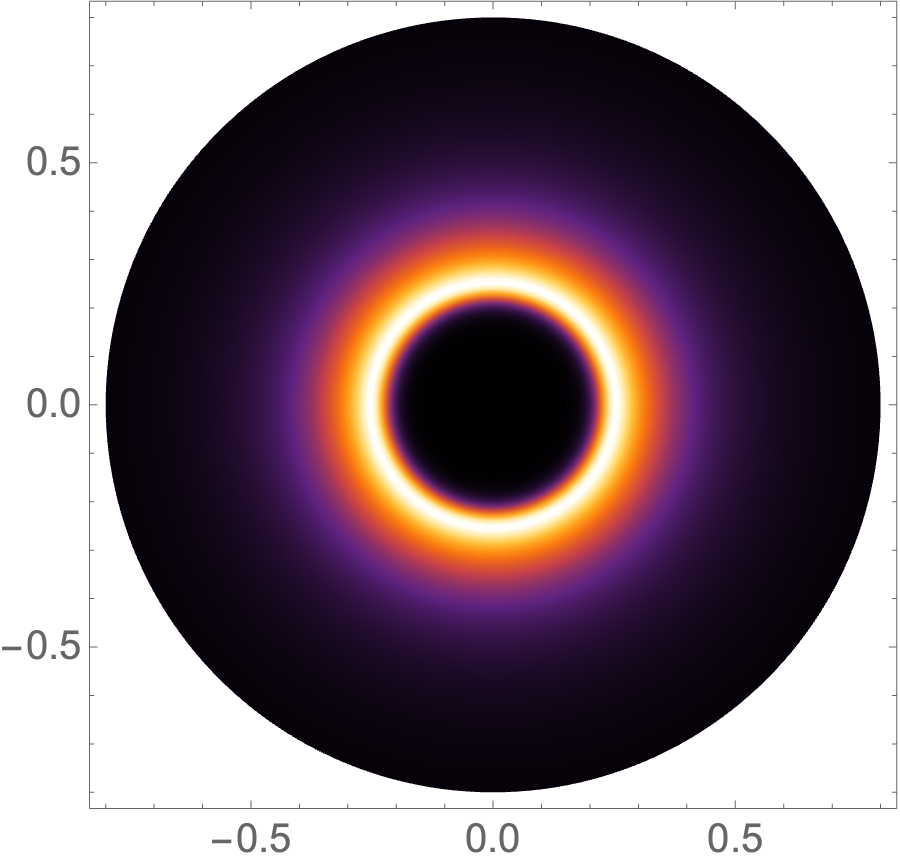}\par\vspace{1mm}
\includegraphics[width=0.30\textwidth]{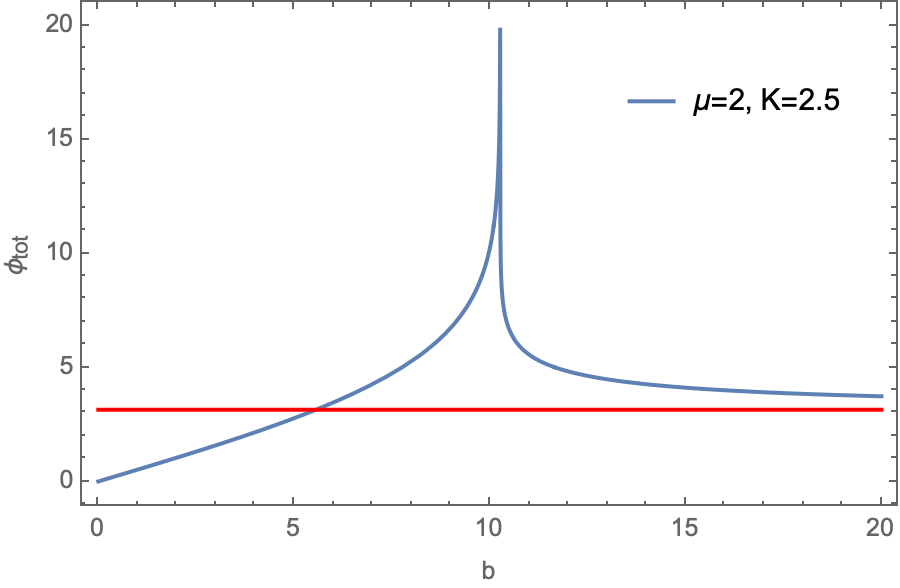}\hspace{0.025\textwidth}%
\includegraphics[width=0.30\textwidth]{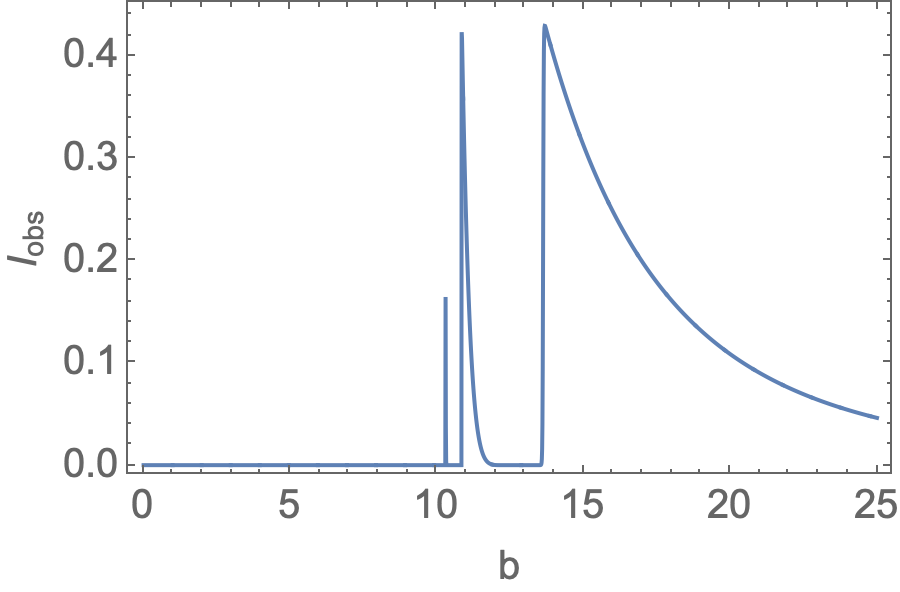}\hspace{0.025\textwidth}%
\includegraphics[width=0.30\textwidth]{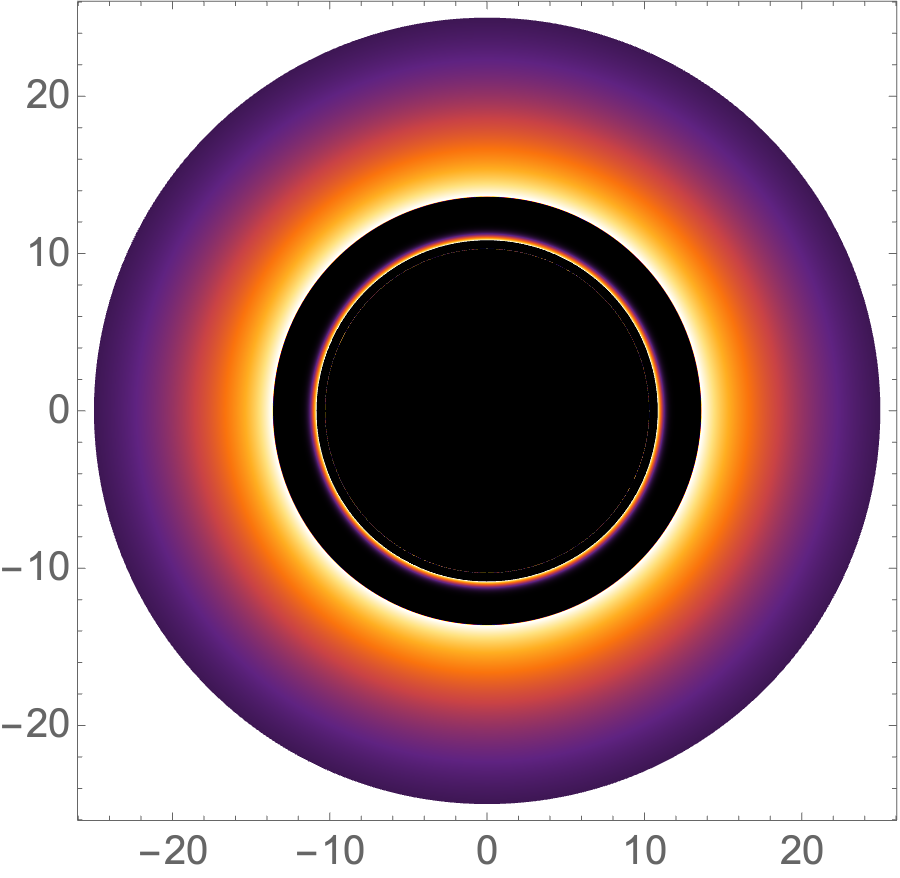}\par\vspace{1mm}
\caption{\it Optical images along path A. 
From left to right, the panels give the total deflection angle $\phi_\text{tot}(b)$, the observed intensity $I_\text{obs}(b)$ and the optical image. 
From top to bottom, the rows correspond to the naked singularity with $\mu=0.2$ and $K=2.5$, the black hole with a static sphere $x_{in}=x_s$, $\mu=0.6$ and $K=2.5$, and the black hole with an ISCO, $\mu=2$ and $K=2.5$.}
\label{pathAimage}
\end{figure}

We first consider path A, with $0<K<K_1$. 
In this case the mass evolution contains a naked singularity, a regular soliton at the regular point, and then a single-horizon black hole. 
We choose $K=2.5$ as a representative example.
The results are shown in Fig.~\ref{pathAimage}.

For the naked singularity with $\mu=0.2$, the deflection angle is a smooth function of $b$. 
It grows from the central region, crosses the line $\phi_\text{tot}=\pi$, reaches a broad maximum and then approaches $\pi$ from above for large $b$. 
There is no divergent peak in the first panel which is consistent with the absence of an unstable light ring in this case. 
The observed intensity is then dominated by the first intersection with  the disk. 
The sharp rise of $I_\text{obs}$ occurs at the lensed position of the inner edge of the disk, while the subsequent decrease follows the falloff of the assumed emission profile. 
The optical image consequently contains a broad bright annulus and a central dark region. This central dark region should not be interpreted as a black hole shadow.
It is produced by the absence of disk emission inside $x_{\rm in}=x_s$
and by the fact that the corresponding photon geodesics do not intersect
the emitting part of the disk.
In addition, a small inner bright ring appears inside the main annulus.
This is different from the naked singularity image in the beyond Horndeski model studied in Ref.~\cite{Huang:2024bbs}, where such an inner ring is absent. 
The reason is that here the deflection angle is large enough to map the disk inner edge to an additional small-$b$ branch of the transfer function. 
This inner ring is therefore a secondary image of the inner edge of the  disk, rather than a light ring associated with a divergent deflection angle.

When the mass increases to $\mu=0.6$, the object is already a black hole, but it belongs to the small region where the disk starts from a static sphere rather than from an ISCO. 
This leads to a much more compact image. 
The scale of the impact parameter in the observed intensity is less than unity, in contrast to the naked-singularity case, where the bright annulus is located at values of $b$ of a few.
The first column of Fig.~\ref{pathAimage} shows that $\phi_\text{tot}$ develops a very sharp peak at small $b$. 
The corresponding peak in $I_\text{obs}$ is mapped to the narrow bright ring close to the dark central region. 
Since the disk inner edge is very close to the compact object, the direct and strongly bent contributions overlap more strongly than in the usual ISCO case.

For the larger mass, $\mu=2$, the solution is a black hole with an ISCO.
The disk starts from the ISCO. 
The image now resembles the familiar thin-disk image of a black hole. 
The first sharp structures in $I_\text{obs}$ correspond to highly bent rays close to the light ring, while the broad outer component is the direct image of the accretion disk. 
The two-dimensional image therefore has a dark central region, a thin
bright ring near the inner boundary and a wider outer annulus. 
Compared with the $\mu=0.6$ case, the whole image is pushed to much larger values of the impact parameter, reflecting the outward displacement of the disk inner edge.

\subsection{Optical images along path B}

\begin{figure}[H]
\centering
\includegraphics[width=0.30\textwidth]{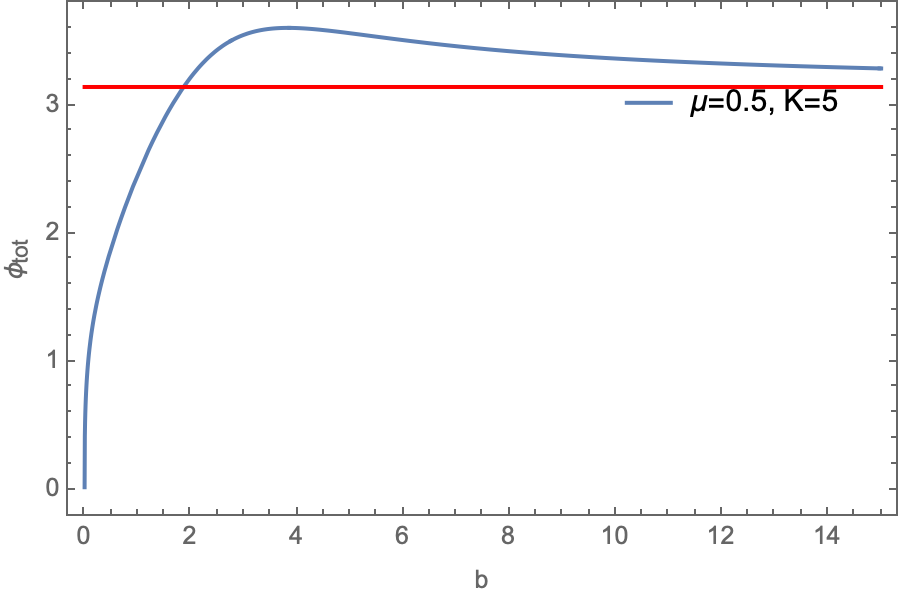}\hspace{0.025\textwidth}%
\includegraphics[width=0.30\textwidth]{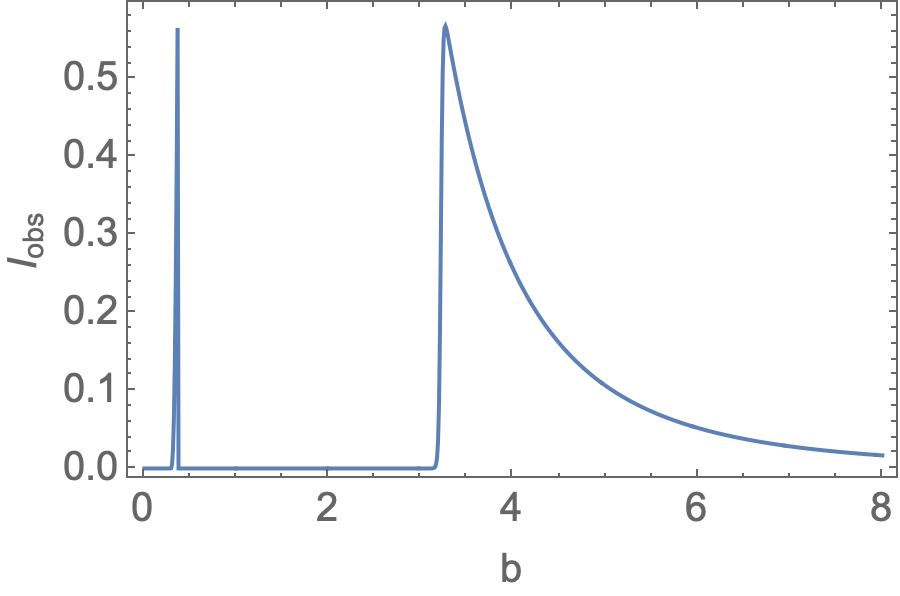}\hspace{0.025\textwidth}%
\includegraphics[width=0.30\textwidth]{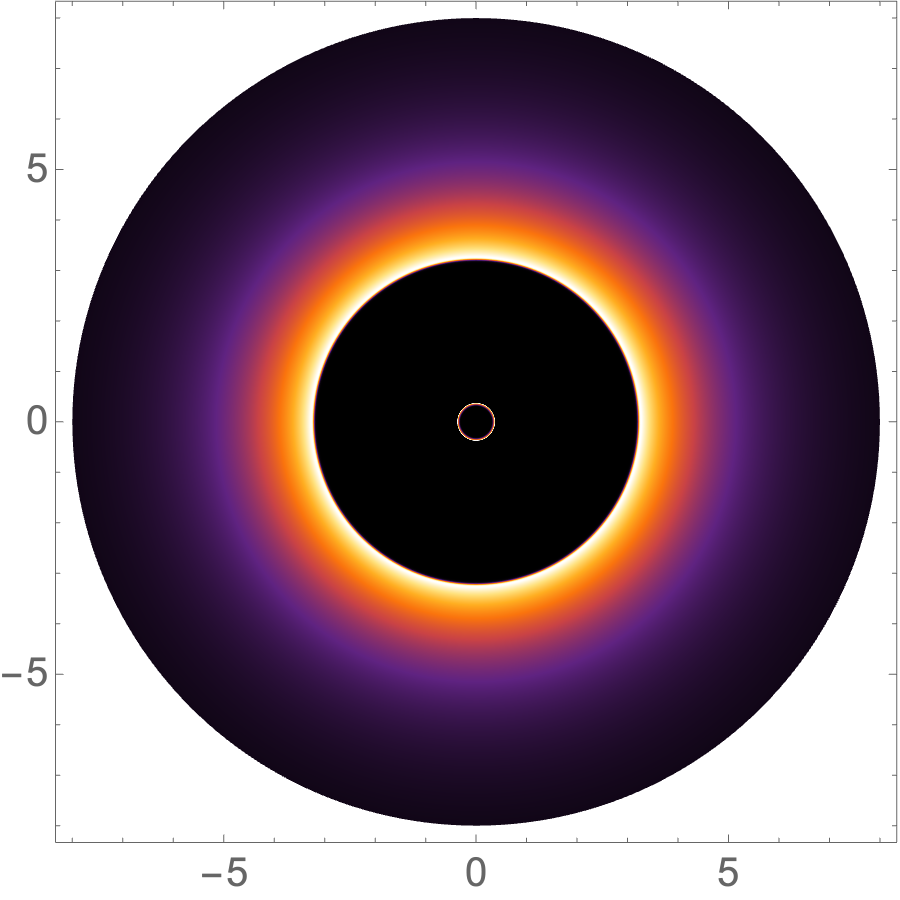}\par\vspace{1mm}
\includegraphics[width=0.30\textwidth]{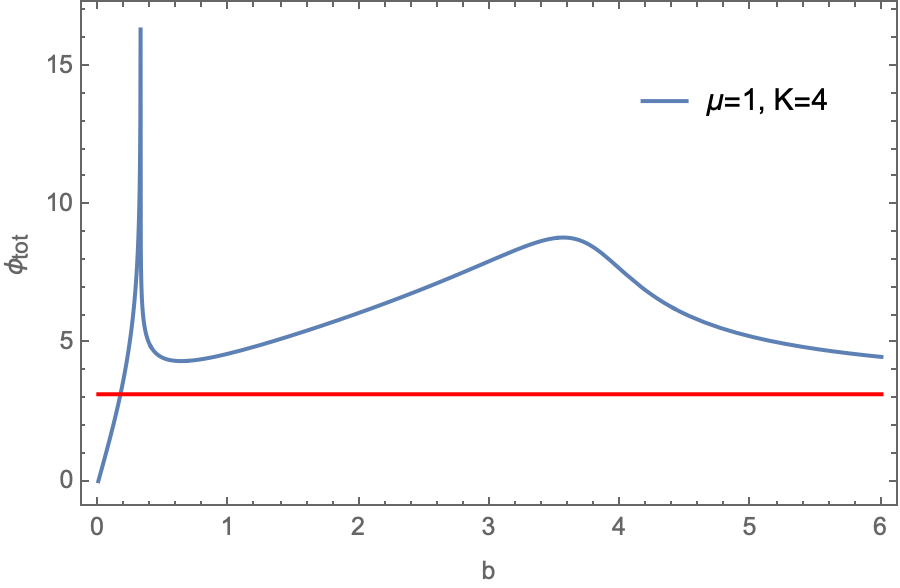}\hspace{0.025\textwidth}%
\includegraphics[width=0.30\textwidth]{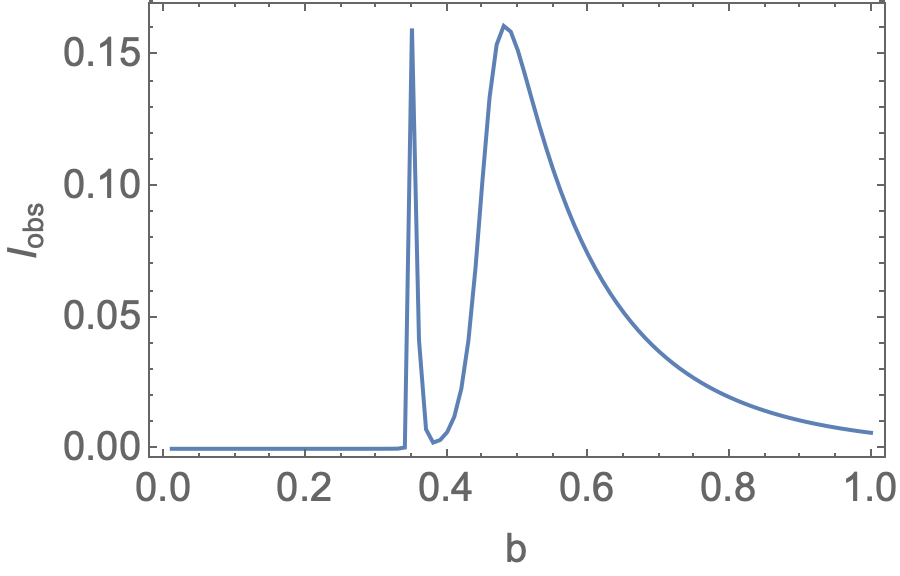}\hspace{0.025\textwidth}%
\includegraphics[width=0.30\textwidth]{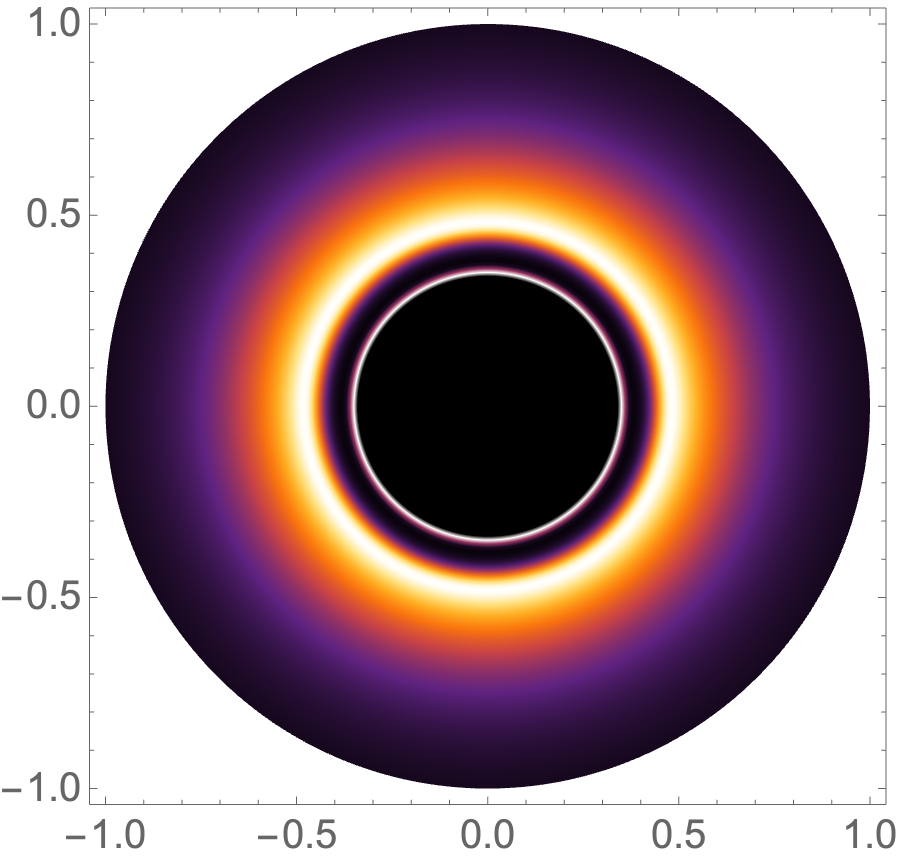}\par\vspace{1mm}
\includegraphics[width=0.30\textwidth]{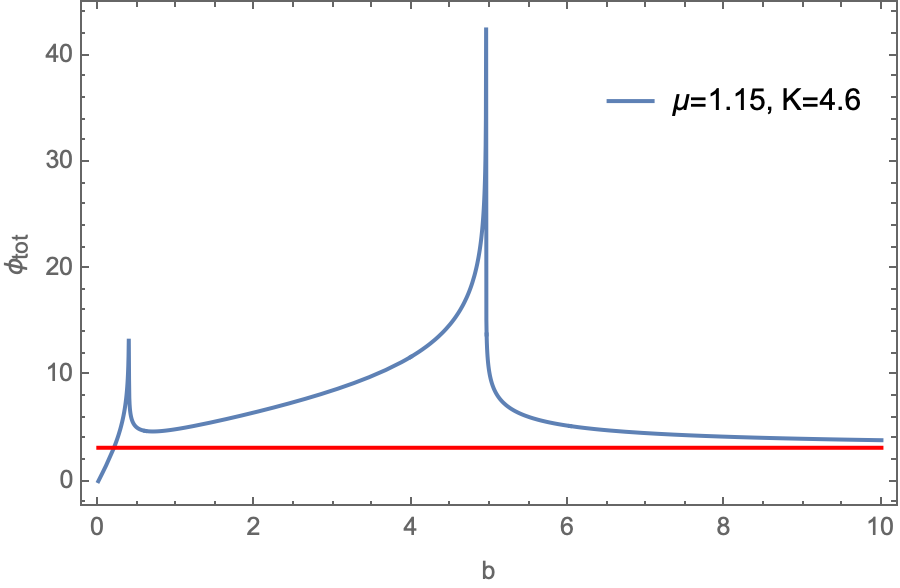}\hspace{0.025\textwidth}%
\includegraphics[width=0.30\textwidth]{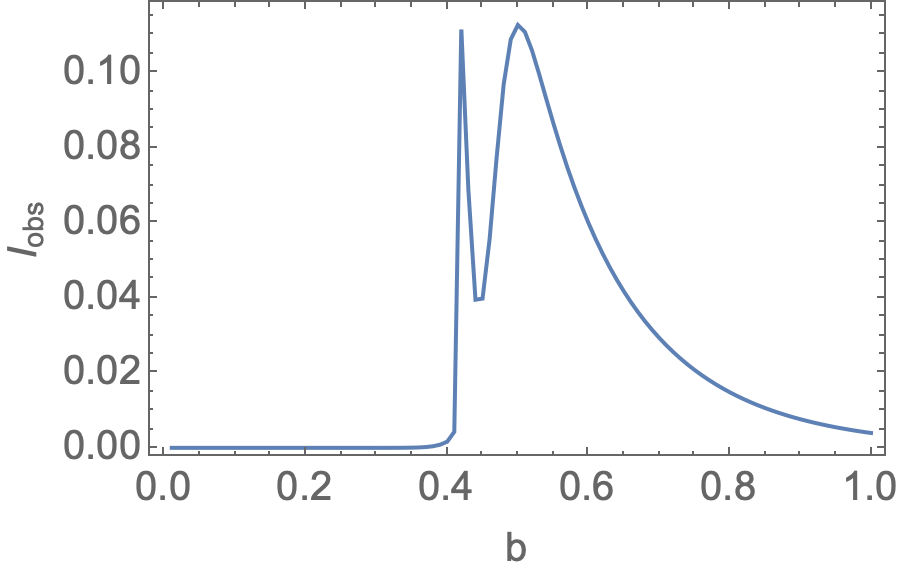}\hspace{0.025\textwidth}%
\includegraphics[width=0.30\textwidth]{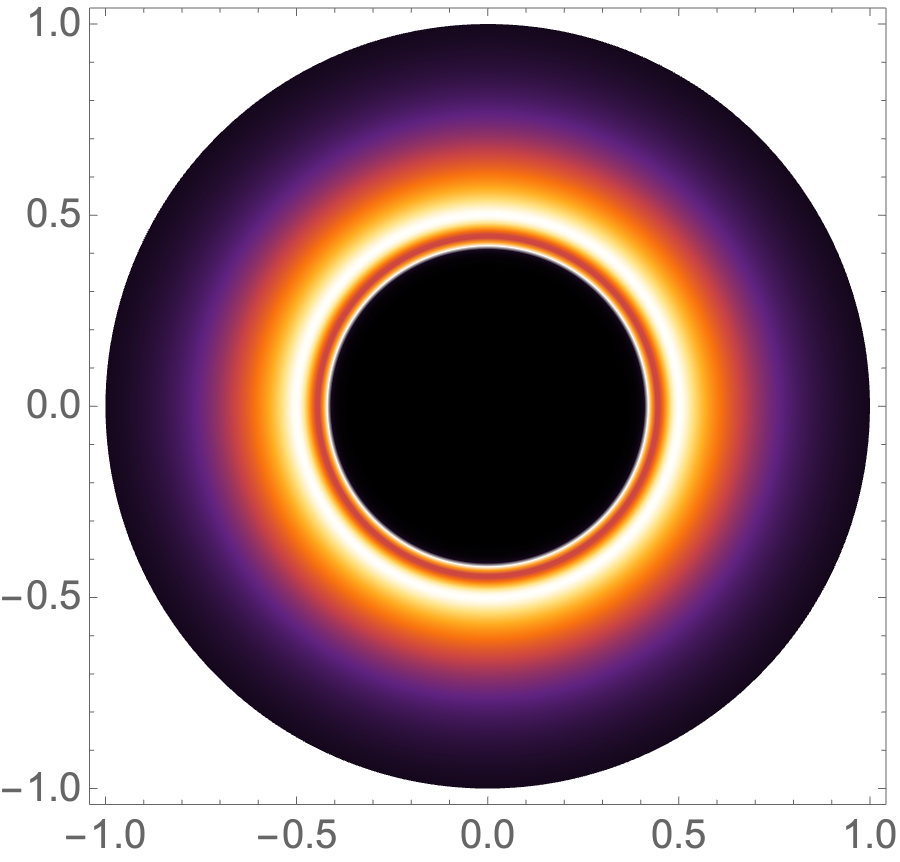}\par\vspace{1mm}
\caption{\it Optical images for representative configurations in path B whose disks start from a static sphere. 
From left to right, the panels give the total deflection angle $\phi_\text{tot}(b)$, the observed intensity $I_\text{obs}(b)$ and the optical image. 
From top to bottom, the rows correspond to the naked singularity with $\mu=0.5$ and $K=5$, the black hole with a static sphere $x_{in}=x_s$ and one light ring, $\mu=1$ and $K=4$, and the black hole with a static sphere $x_{in}=x_s$ and two light rings, $\mu=1.15$ and $K=4.6$.}
\label{pathBimage1}
\end{figure}

\begin{figure}[H]
\centering
\includegraphics[width=0.30\textwidth]{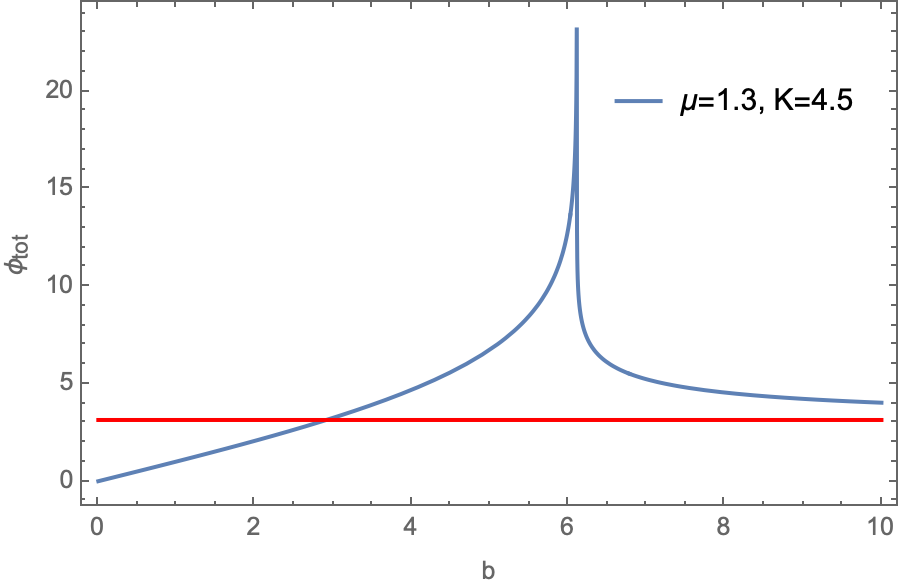}\hspace{0.025\textwidth}%
\includegraphics[width=0.30\textwidth]{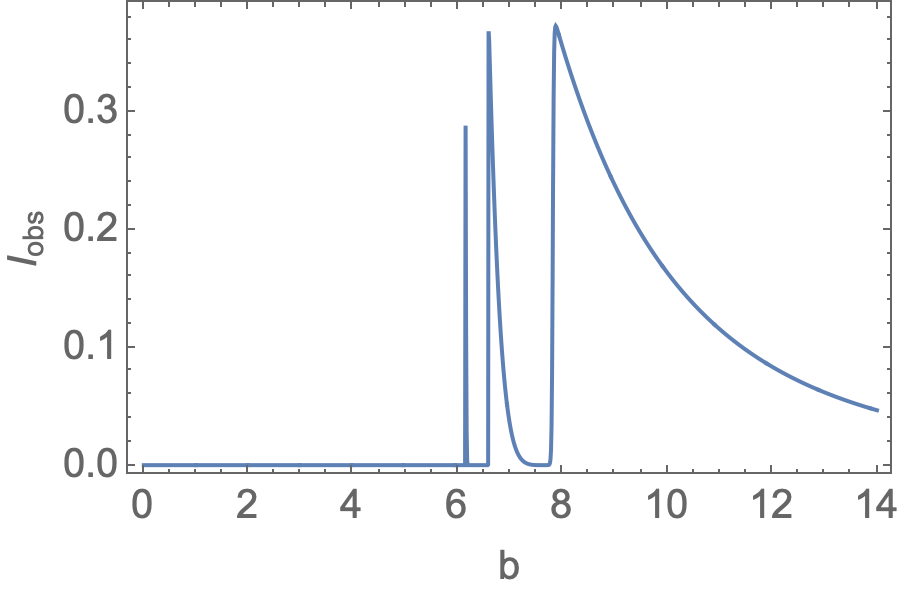}\hspace{0.025\textwidth}%
\includegraphics[width=0.30\textwidth]{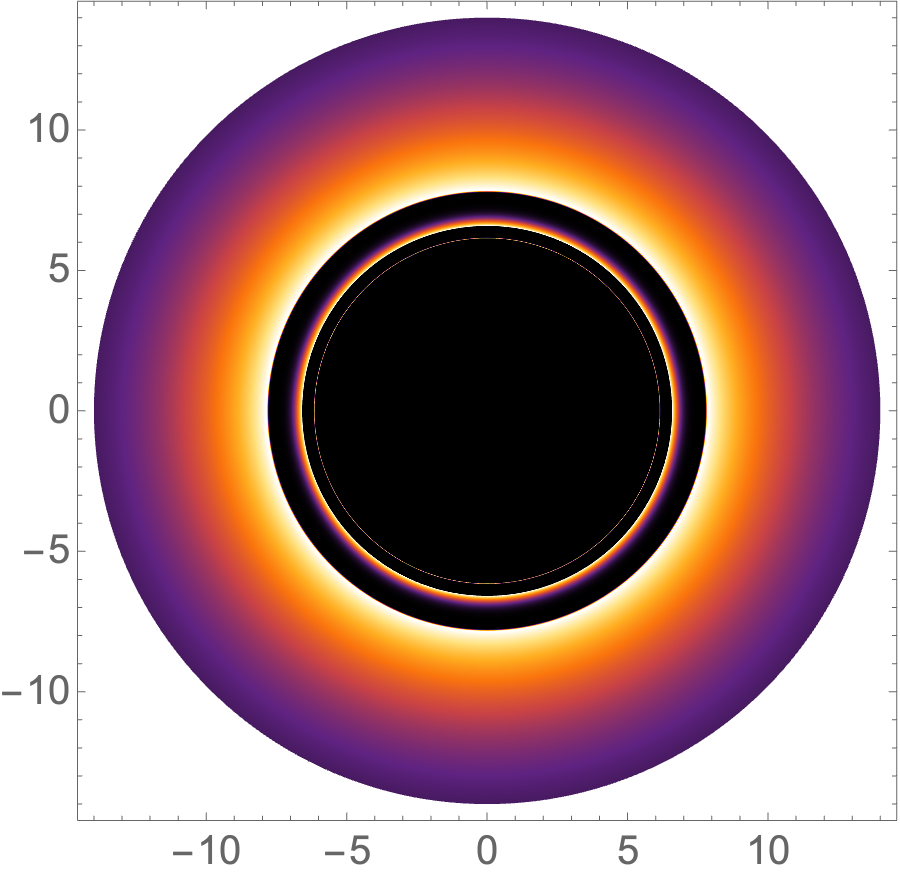}\par\vspace{1mm}
\includegraphics[width=0.30\textwidth]{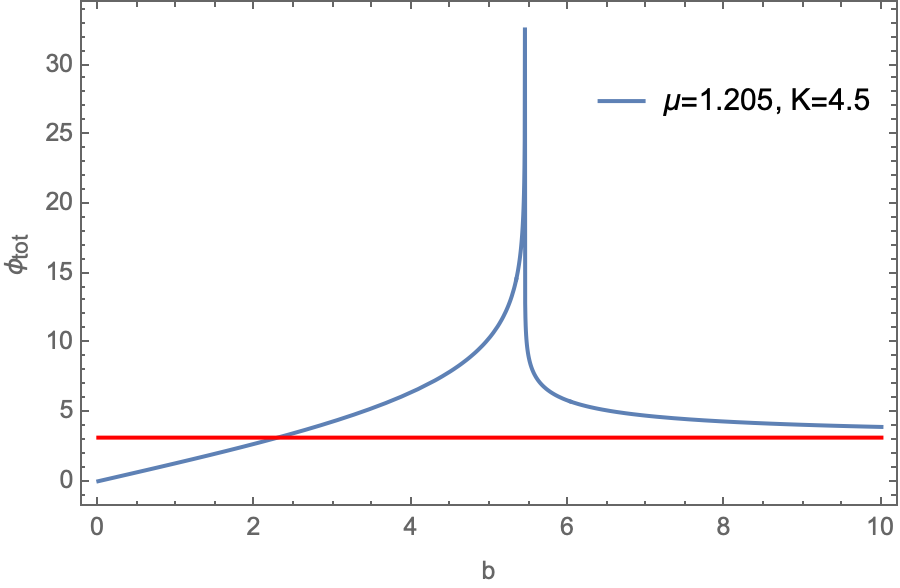}\hspace{0.025\textwidth}%
\includegraphics[width=0.30\textwidth]{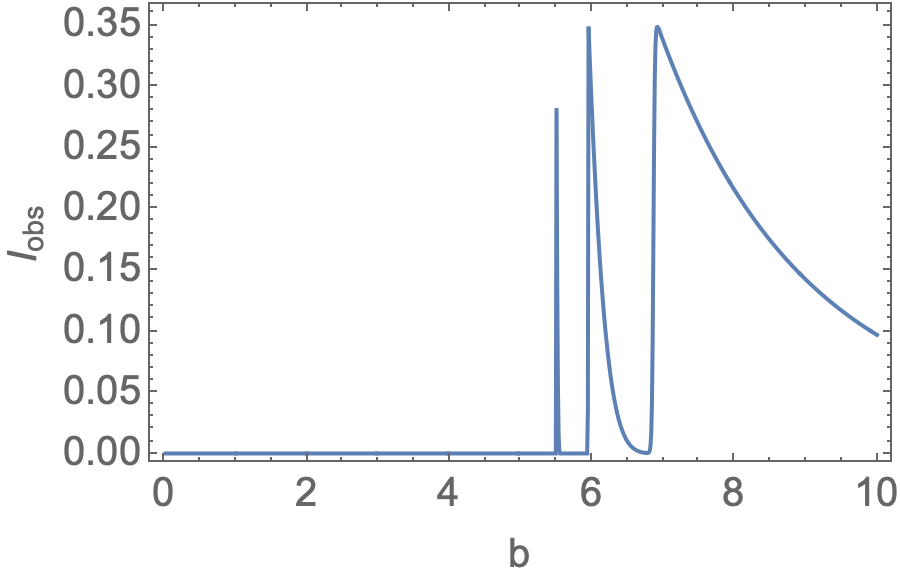}\hspace{0.025\textwidth}%
\includegraphics[width=0.30\textwidth]{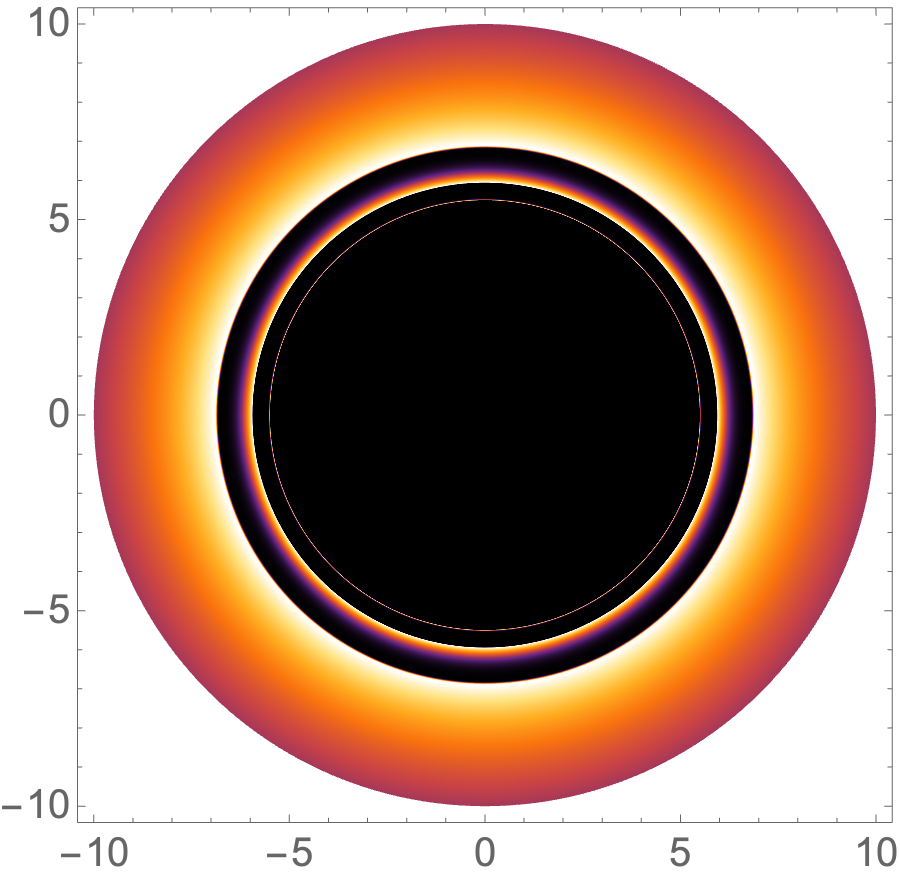}\par\vspace{1mm}
\caption{\it Optical images for the black hole configurations with ISCO in path B. 
From left to right, the panels give the total deflection angle $\phi_\text{tot}(b)$, the observed intensity $I_\text{obs}(b)$ and the optical image.
From top to bottom, the rows correspond to the single-horizon black hole with an ISCO, $\mu=1.3$ and $K=4.5$, and the three-horizon black hole with an ISCO, $\mu=1.205$ and $K=4.5$. 
The three-horizon black hole has an optical appearance close to the single-horizon case, because the image is mainly controlled by the outer light ring and the disk inner edge.}\label{pathBimage2}
\end{figure}

For path B, $K_1<K<K_2$, the solution space is richer. 
In addition to the naked singularity and the single-horizon black holes, there is a finite interval where three horizons are present. 
We show several representative cases in Figs.~\ref{pathBimage1} and \ref{pathBimage2}.

The first row of Fig.~\ref{pathBimage1} corresponds to a naked singularity with $\mu=0.5$ and $K=5$. 
The behavior is already different from the naked singularity in path A.
The deflection angle becomes larger than $\pi$ for a wide range of $b$, which means that some light rays can bend enough to form secondary images of the disk. 
However, the curve does not show the same logarithmic divergence as in the black-hole light ring cases shown below.
Accordingly, the observed intensity contains a dominant peak associated with the disk inner edge and then decreases smoothly. 
The resulting optical image has a bright annulus surrounding a central dark region. 
Again this dark region is not produced by a horizon, but by the combination of the static-sphere cutoff of the disk and the transfer function. 
The optical image also contains an inner bright ring. 
In the present case it is produced by the additional lensed image of the disk inner edge, because the strong deflection of the light rays makes the transfer function nontrivial even in the absence of a horizon.

The second row of Fig.~\ref{pathBimage1} shows a black hole with $\mu=1$ and $K=4$, for which the accretion disk starts from a static sphere.
The image is much smaller than the ISCO images, because the inner edge of the disk is close to the central object. 
The intensity profile displays a narrow feature at small impact parameter followed by a broader peak. 
These two features have different origins. 
The narrow feature comes from highly bent light rays close to the critical curve, while the broader peak is associated with the direct image of the disk. 
This separation is also visible in the image, where a thin bright ring appears inside a wider luminous annulus.

The third row, $\mu=1.15$ and $K=4.6$, is one of the most characteristic
examples in this family. 
The null effective potential has two unstable light ring maxima separated by a minimum. 
Correspondingly, $\phi_\text{tot}(b)$ has two peaks. 
The first one is associated with the inner scattering branch, while the second one is associated with the outer light ring. 
This multi-branch structure is inherited by $I_\text{obs}(b)$, where the intensity has more than one narrow feature before the broad direct-emission component. 
In the optical image, the effect is a visible fine structure close to the inner dark region. 
Thus this object does not simply mimic a Schwarzschild-like black hole. 
The existence of two light ring scales leaves an imprint in the radial brightness profile.

When the disk starts from the ISCO, the optical image becomes more similar to the standard black-hole case. 
We show this situation in Fig.~\ref{pathBimage2}. 
The first row is a single-horizon black hole with $\mu=1.3$ and $K=4.5$.
The deflection angle diverges at the critical impact parameter of the light ring. 
The observed intensity has a set of narrow peaks near the shadow edge, followed by the main broad peak associated with direct disk emission. 
These narrow peaks are the lensing and higher-order ring contributions. 
The dark gap between the inner ring and the broad outer emission is a consequence of the transfer function, because in that interval of $b$, the rays do not intersect the bright part of the disk efficiently.

The second row of Fig.~\ref{pathBimage2} is a three-horizon black hole with $\mu=1.205$ and $K=4.5$. 
Its image is remarkably close to the single-horizon case. 
This is physically reasonable. 
The observer at infinity is mainly sensitive to the outermost light ring, the outer horizon and the position of the ISCO. 
The additional inner horizons are hidden behind the outer trapping region and do not directly determine the transfer function of the photons that reach the observer. 
The multi-horizon nature of the spacetime is therefore imprinted only indirectly, through its effect on the location of the outer light ring and on the bound timelike orbits.

\subsection{Optical images along path C}

\begin{figure}[H]
\centering
\includegraphics[width=0.30\textwidth]{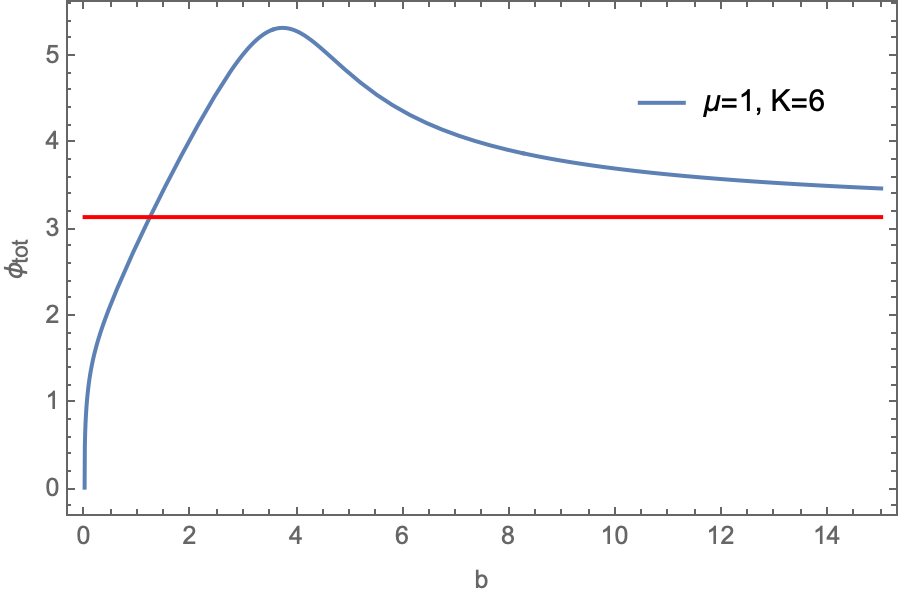}\hspace{0.025\textwidth}%
\includegraphics[width=0.30\textwidth]{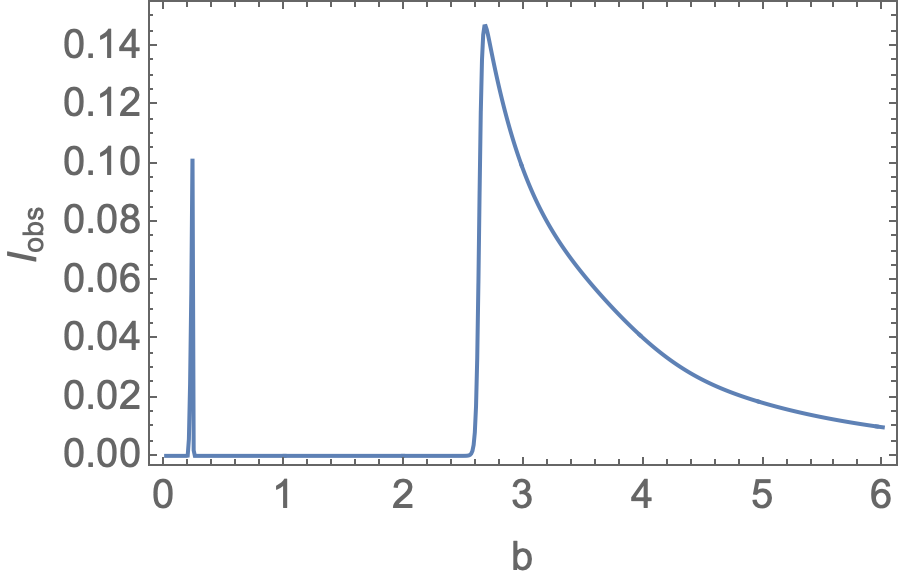}\hspace{0.025\textwidth}%
\includegraphics[width=0.30\textwidth]{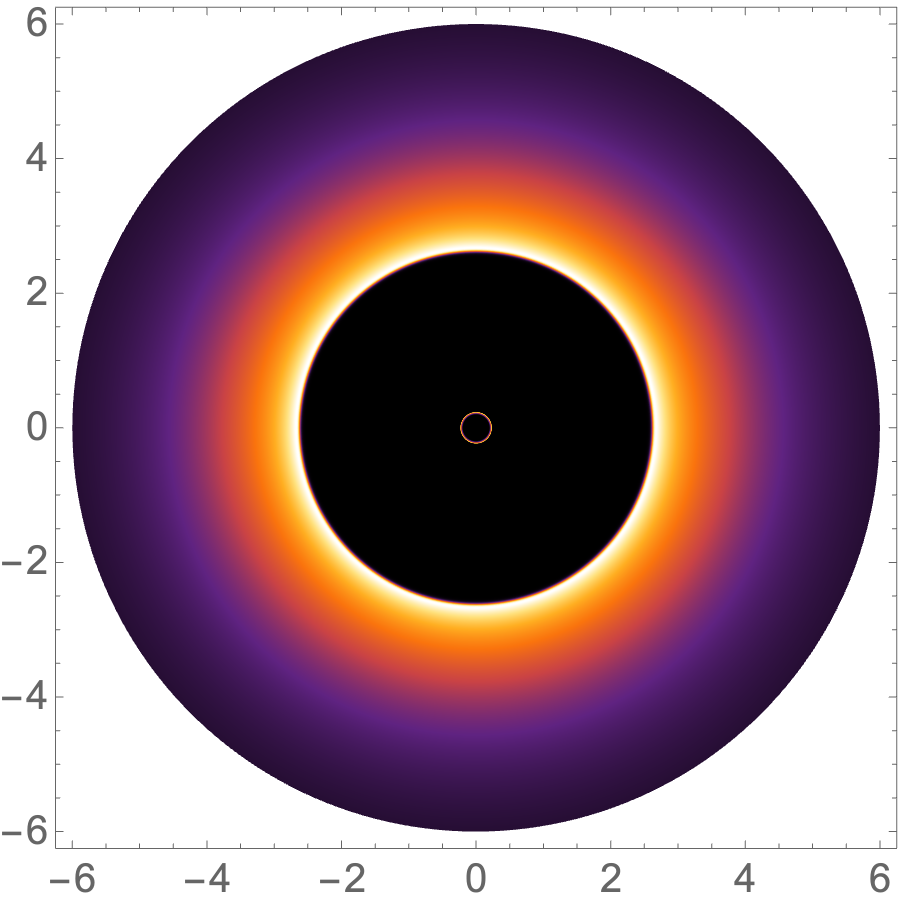}\par\vspace{1mm}
\includegraphics[width=0.30\textwidth]{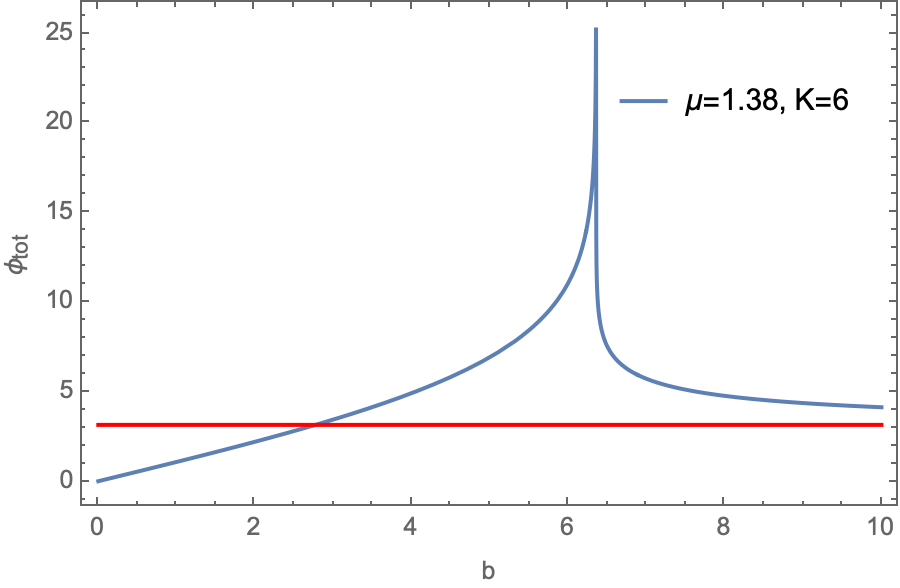}\hspace{0.025\textwidth}%
\includegraphics[width=0.30\textwidth]{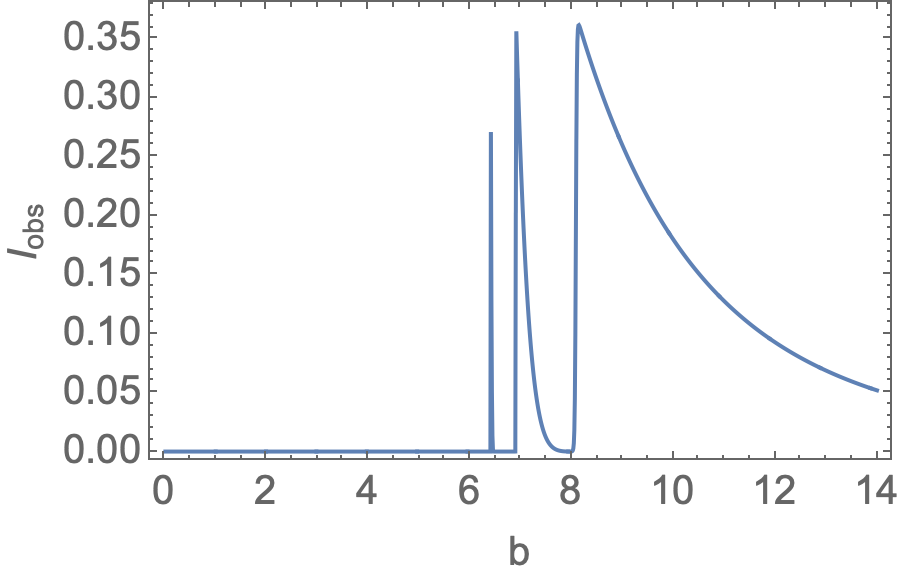}\hspace{0.025\textwidth}%
\includegraphics[width=0.30\textwidth]{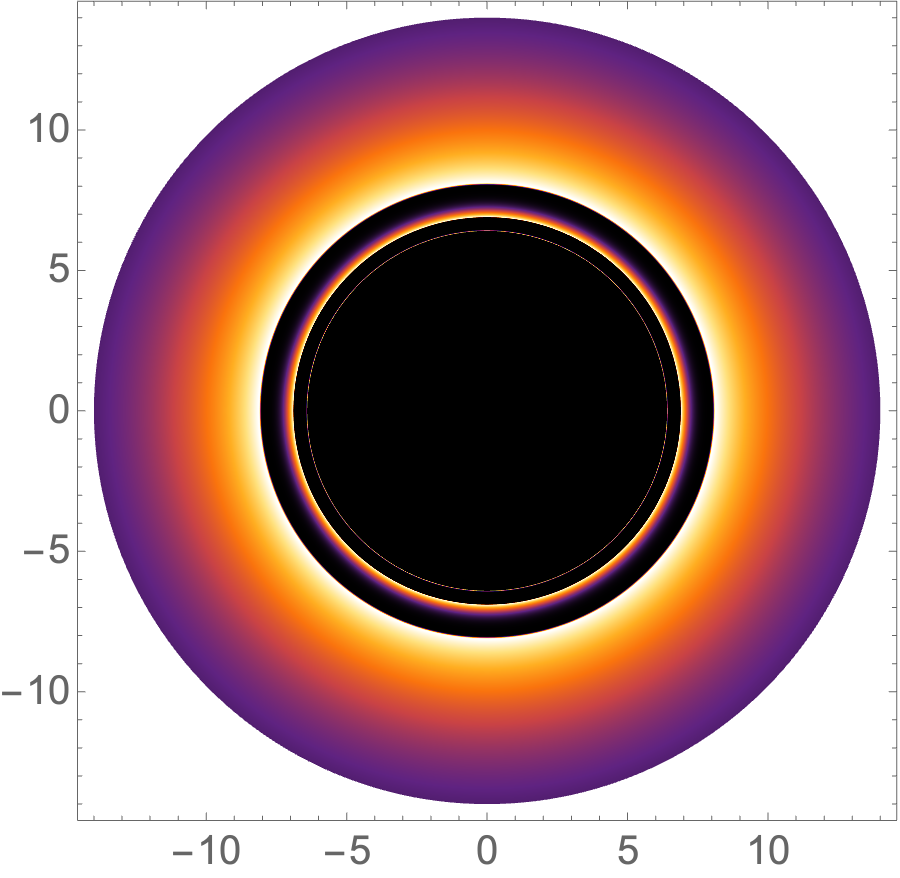}\par\vspace{1mm}
\caption{\it Optical images along path C before and after the formation of horizons. 
From left to right, the panels give the total deflection angle $\phi_\text{tot}(b)$, the observed intensity $I_\text{obs}(b)$ and the optical image.
From top to bottom, the rows correspond to the naked singularity with
$\mu=1$ and $K=6$, and the two-horizon black hole with $\mu=1.38$ and $K=6$.
The naked singularity already exhibits strong bending, while the two-horizon black hole has a genuine shadow.}\label{pathCimage1}
\end{figure}

\begin{figure}[H]
\centering
\includegraphics[width=0.30\textwidth]{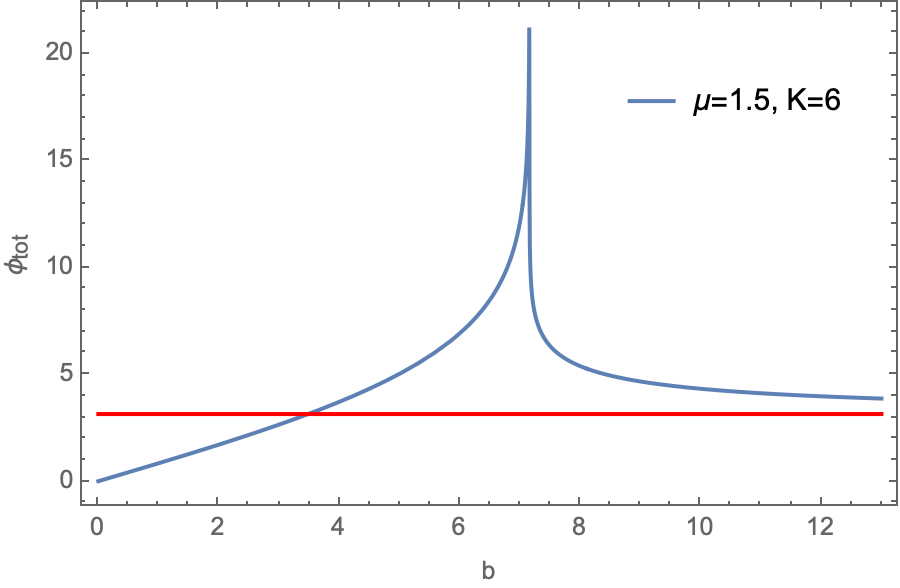}\hspace{0.025\textwidth}%
\includegraphics[width=0.30\textwidth]{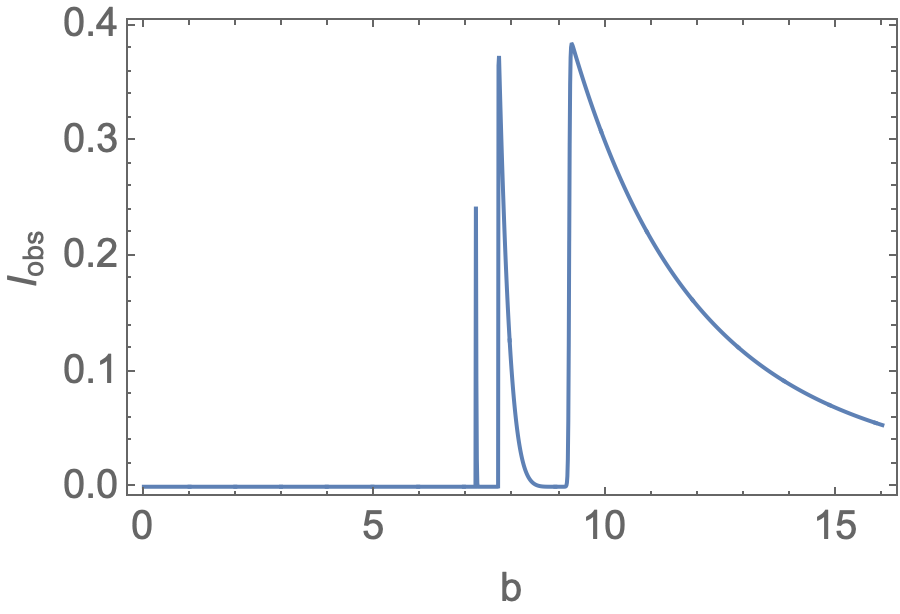}\hspace{0.025\textwidth}%
\includegraphics[width=0.30\textwidth]{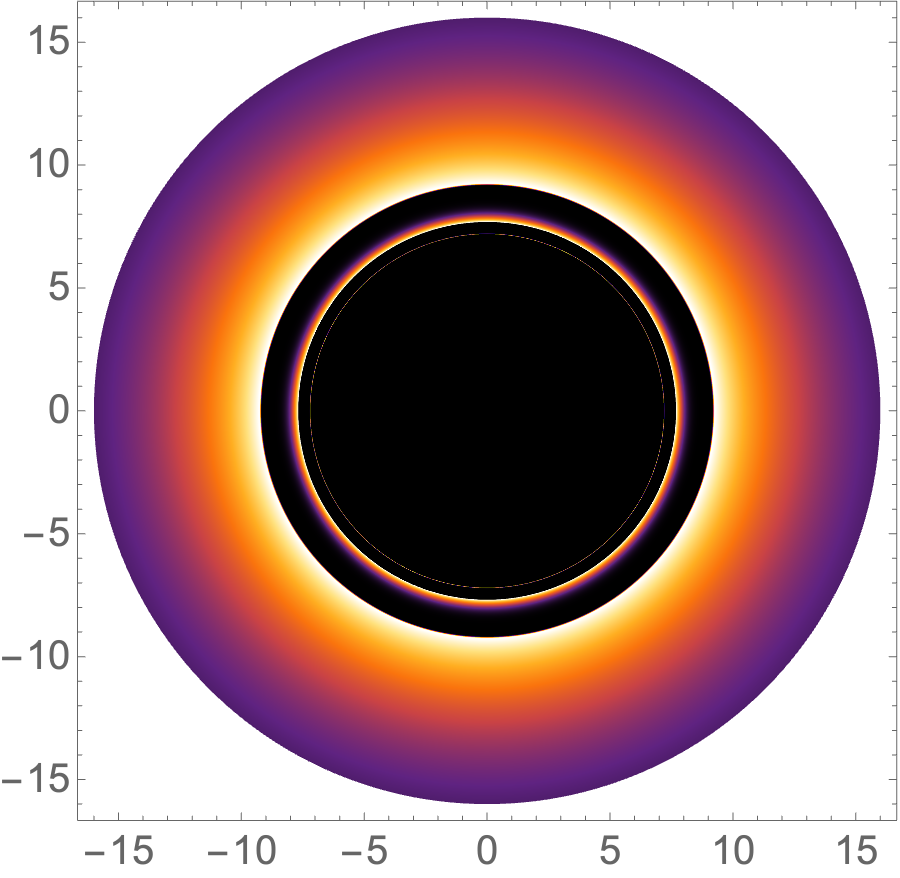}\par\vspace{1mm}
\includegraphics[width=0.30\textwidth]{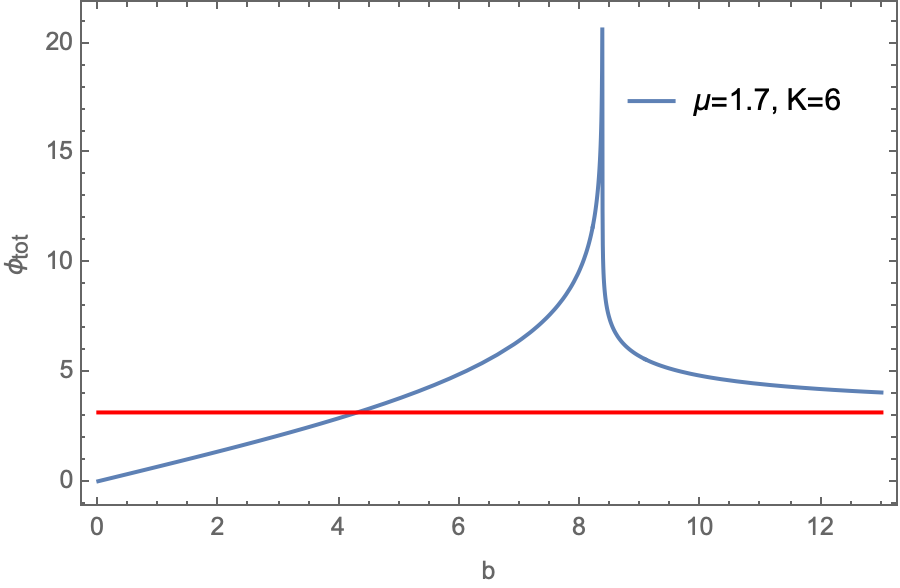}\hspace{0.025\textwidth}%
\includegraphics[width=0.30\textwidth]{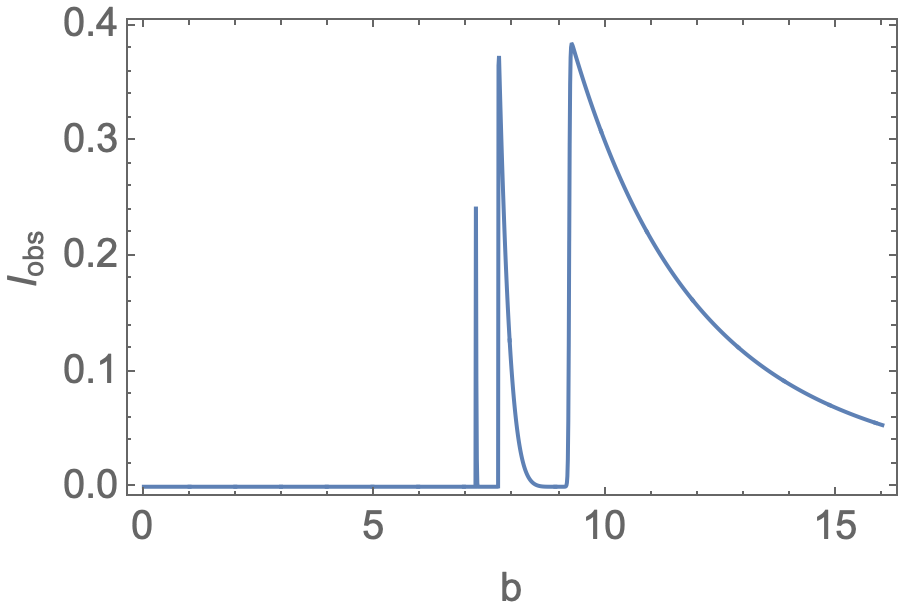}\hspace{0.025\textwidth}%
\includegraphics[width=0.30\textwidth]{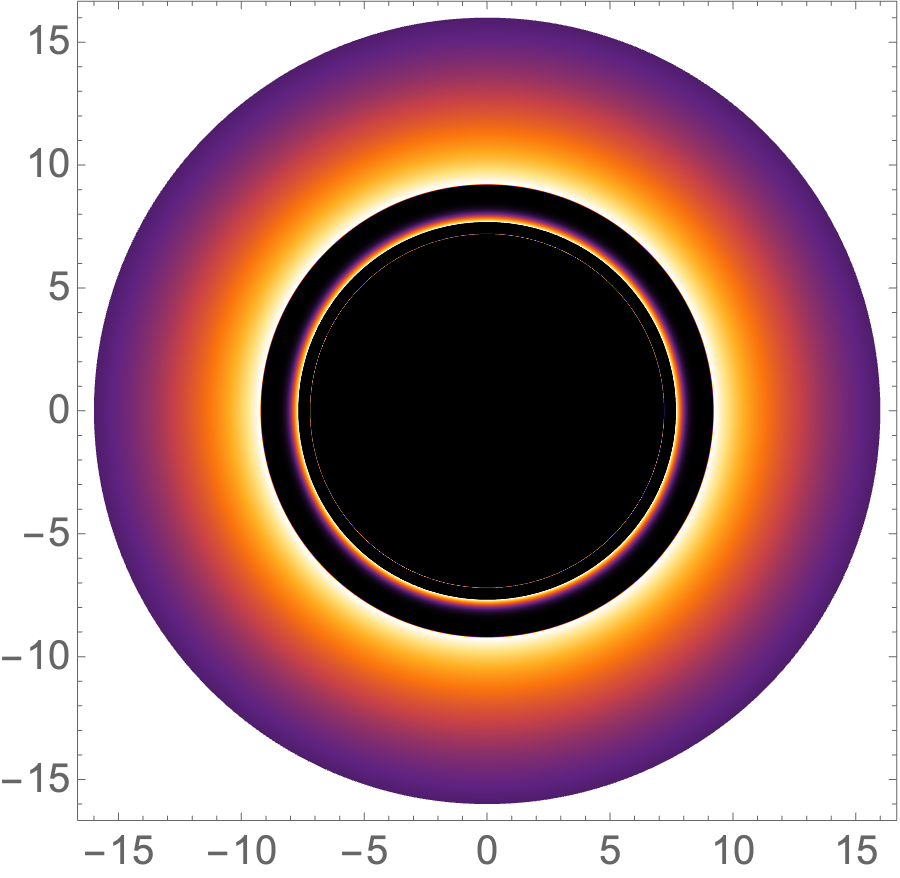}\par\vspace{1mm}
\caption{\it Optical images for the black-hole configurations with ISCO in path C. 
From left to right, the panels give the total deflection angle $\phi_\text{tot}(b)$, the observed intensity $I_\text{obs}(b)$ and the optical image.
From top to bottom, the rows correspond to the three-horizon black hole with an ISCO, $\mu=1.5$ and $K=6$, and the single-horizon black hole with an ISCO, $\mu=1.7$ and $K=6$. 
The images are mainly controlled by the outer light ring and by the ISCO, while the inner horizon structure leaves only an indirect imprint.}
\label{pathCimage2}
\end{figure}

For path C, $K>K_2$, the regular point lies inside the black-hole region.
We choose $K=6$ as a representative example. 
As discussed in section 3, the mass path contains a naked singularity, a two-horizon black hole, a regular black hole, a three-horizon black hole and finally a single-horizon black hole.
The corresponding images are shown in Figs.~\ref{pathCimage1} and \ref{pathCimage2}.

In contrast to path A, the naked singularity in path C already exhibits strong light bending. 
In the first row of Fig.~\ref{pathCimage1}, the curve $\phi_\text{tot}(b)$ has a broad maximum well above $\pi$. 
Consequently, the observed intensity has a pronounced peak at the lensed disk edge and a long tail at larger impact parameters. 
The image is therefore more compact than the naked singularity images in paths A and B, and it contains a sharper bright annulus. 
The important point is that the image already looks shadow-like before the formation of any horizon. 
This again shows that the central depression in a thin-disk image is not, by itself, a proof of the existence of an event horizon. 
The inner bright ring in this naked singularity case has the same origin as in paths A and B. 

The second row of Fig.~\ref{pathCimage1} shows the two-horizon black hole
with $\mu=1.38$ and $K=6$. 
Once horizons appear, the central dark region has a genuine capture interpretation, since photons with sufficiently small impact parameters fall through the outer horizon instead of returning to infinity.
At the same time, the divergent behavior of $\phi_\text{tot}$ at the light ring produces narrow peaks in $I_\text{obs}$. 
These peaks are mapped to the thin bright rings close to the boundary of the shadow. 
Compared with the naked singularity in the first row, the radius of the main bright annulus is larger and the separation between the central dark region and the outer emission becomes clearer.

For larger values of the mass, the black holes with an ISCO display the
standard hierarchy of optical features, including a central dark region, a narrow lensing ring and an extended outer emission region.
This is shown in Fig.~\ref{pathCimage2}. 
The first row is a three-horizon black hole with $\mu=1.5$, while the second row is a single-horizon black hole with $\mu=1.7$.
The two images are extremely similar. 
In both cases, the dominant direct emission forms the broad outer annulus, and the narrow peaks in the observed intensity form the thin rings close to the shadow edge. 
The increase of $\mu$ shifts the relevant optical structures to larger impact parameters, but it does not create a qualitatively new feature in the image.

This comparison makes clear that the number of horizons is not directly visible in the image. 
A three-horizon black hole and a single-horizon black hole can have almost the same optical appearance if their exterior light ring and ISCO structures are similar. 
The inner horizons are hidden from the observer in the sense of ray tracing; photons that contribute to the observed image are controlled by the exterior potential barrier and by the disk emission region.

\subsection{Quantitative image diagnostics and radial-profile comparisons}

We first summarize, for the representative configurations shown in Figs.~\ref{pathAimage} - \ref{pathCimage2}, the geometric quantities that determine the optical images. These are listed in Table~\ref{tab1} \cite{Mizuno:2018lxz,Johnson:2019ljv,Vincent:2020dij,Gralla:2020srx,Younsi:2021dxe,Broderick:2022tfu}. 
Besides the number of horizons $N_{\rm H}$, we give the number of unstable light rings relevant for outer photon propagation $N_{\rm LR}$, the outermost unstable light-ring radius $x_{\rm LR}^{\rm out}$, the corresponding critical impact parameter $b_{\rm crit}$, and the inner edge of the disk $x_{\rm in}$. 
Here $x_{\rm in}=x_{\rm ISCO}$ for disks starting at the ISCO, while $x_{\rm in}=x_s$ for disks starting at a static sphere.

\begin{table}[h]
\centering
\caption{\it Image diagnostic quantities for representative configurations. Here $x_{\rm in}=x_{\rm ISCO}$ for disks starting at the ISCO and $x_{\rm in}=x_s$ for disks starting at a static sphere.\vspace{0.2cm}}
\begin{tabular}{cccccccccc}
\toprule
Path & $K$ & $\mu$ & Object type & $N_{\rm H}$ & $N_{\rm LR}$ & $x_{\rm LR}^{\rm out}$ & $b_{\rm crit}$ & $x_{\rm in}$ & $b_{\rm peak}$  \\
\midrule
A & $2.5$ & $0.2$ & naked singularity & 0 & 0 & -- & -- & $x_s$ & 3.505  \\
A & $2.5$ & $0.6$ & black hole, static sphere & 1 & 1 & 0.032 & 0.056 & $x_s$ & 0.253  \\
A & $2.5$ & $2.0$ & black hole, ISCO & 1 & 1 & 5.878 & 10.265 & $x_{\rm ISCO}$ & 13.684  \\
B & $4.5$ & $1.205$ & three-horizon black hole & 3 & 1 & 2.781 & 5.448 & $x_{\rm ISCO}$ & 6.916  \\
B & $4.5$ & $1.3$ & single-horizon black hole & 1 & 1 & 3.259 & 6.111 & $x_{\rm ISCO}$ & 7.870  \\
C & $6.0$ & $1.5$ & three-horizon black hole & 3 & 1 & 3.866 & 7.158 & $x_{\rm ISCO}$ & 9.275  \\
C & $6.0$ & $1.7$ & single-horizon black hole & 1 & 1 & 4.646 & 8.369 & $x_{\rm ISCO}$ & 10.975  \\
\bottomrule
\label{tab1}
\end{tabular}
\end{table}

The quantities in Table~\ref{tab1} already illustrate the main physical
point. Configurations with different horizon structure can have similar exterior optical scales, while configurations belonging to the same broad class can have different image sizes because their light-ring radii, critical impact parameters or disk inner edges differ. 
Thus the number of horizons alone is not sufficient to characterize the image; the 
photon potential and the disk inner edge provide the more direct diagnostics.

\begin{table}[h]
\centering
\caption{\it Symmetric $L^1$ comparisons between the single-horizon and three-horizon black holes in paths B and C. 
The column $\Delta^{\rm sym}_{L^1}$ compares the profiles at fixed impact parameter, while $\Delta^{\rm shape,sym}_{L^1}$  compares them after rescaling $b$ by the corresponding critical impact parameter $b_{\rm crit}$.}
\vspace{0.2cm}
\begin{tabular}{cccccccc}
\toprule
Path & $K$ & $\mu_1$ & $\mu_2$ & $\Delta_{L^1}^{\rm sym}$ & $\Delta_{L^1}^{\rm shape,sym}$ \\
\midrule
B & $4.5$ & $1.205$ & $1.3$ &  0.38587&  0.09084\\
C & $6.0$ & $1.5$   & $1.7$ &  0.46732&  0.07386\\
\bottomrule
\label{tab2}
\end{tabular}
\end{table}

\begin{table}[h]
\centering
\caption{\it Symmetric NRMSE-type fixed-scale comparisons between the single-horizon and three-horizon black holes in paths B and C. 
The profiles are compared at the same impact parameter $b$.
\vspace{0.2cm}}
\begin{tabular}{cccccccc}
\toprule
Path & $K$ & $\mu_1$ & $\mu_2$ & ${\rm NRMSE}^{\rm sym}_{2D}$  & ${\rm NRMSE}_{\rm profile}^{\rm sym}$ \\
\midrule
B & $4.5$ & $1.205$ & $1.3$ &  0.55940  & 0.59647\\
C & $6.0$ & $1.5$   & $1.7$ &  0.61926  & 0.65323 \\
\bottomrule
\label{tab3}
\end{tabular}
\end{table}

We next compare the radial intensity profiles themselves. 
This comparison is important because two images may look qualitatively similar while their bright rings and direct-emission peaks occur at different impact parameters.
For two images with radial intensity profiles $I_1(b)$ and $I_2(b)$, we define the symmetrized normalized pointwise $L^1$ distance
\begin{equation}
\Delta_{L^1}^{\rm sym}
=
\frac{\displaystyle \int_{b_{\rm min}}^{b_{\rm max}}
\left| I_1(b)-I_2(b) \right|\, db}
{\displaystyle \int_{b_{\rm min}}^{b_{\rm max}} (I_1(b)+I_2(b))\, db}
\, .
\end{equation}
Here $b_{\min}$ and $b_{\max}$ denote the lower and upper limits of the impact-parameter interval used in the comparison. 
This quantity compares the two profiles at the same value of $b$, and is therefore sensitive to shifts of the whole image scale.

To separate such scale changes from genuine differences in morphology, we also introduce a scale-free comparison. 
For each image we define
\be
\hat b_i = \frac{b}{b_{{\rm crit},i}},
\ee
where $b_{{\rm crit},i}$ is the critical impact parameter of the outer unstable light ring of the $i$-th spacetime. 
The corresponding rescaled profile is
\be
\tilde I_i(\hat b_i) = I_i(b_{{\rm crit},i}\hat b_i).
\ee
We then define
\be
\Delta_{L^1}^{\rm shape,sym}
=
\frac{
\displaystyle
\int_{\hat b_{\min}}^{\hat b_{\max}}
\left|
\tilde I_1(\hat b)-\tilde I_2(\hat b)
\right|\,d\hat b
}{
\displaystyle
\int_{\hat b_{\min}}^{\hat b_{\max}}
(\tilde I_1(\hat b)+\tilde I_2(\hat b))\,d\hat b
}.
\ee
This quantity compares the shape of the two radial profiles after the critical-curve scale has been factored out. 

As complementary diagnostics we also compute symmetrized NRMSE-type measures, adapted from the image-comparison measures used in Refs.~\cite{Chael:2018oym,Emami:2022ydq}.
The standard NRMSE definitions are usually reference-normalized and therefore non-symmetric; here we use symmetric normalizations because neither of the two beyond-Horndeski configurations is taken as a preferred reference. The area-weighted version,
\be
{\rm NRMSE}_{2D}^{\rm sym}
=
\left[
\frac{
\displaystyle
\int_{b_{\min}}^{b_{\max}}
\left(I_1(b)-I_2(b)\right)^2\,b\,db
}{
\displaystyle
\int_{b_{\min}}^{b_{\max}} ( I_1(b)^2
+ I_2(b)^2\, )
\,b\,db
}
\right]^{1/2}.
\ee
compares the corresponding two-dimensional circular images, whereas 
\be
{\rm NRMSE}_{\rm profile}^{\rm sym}
=
\left[
\frac{
\displaystyle
\int_{b_{\min}}^{b_{\max}}
\left(I_1(b)-I_2(b)\right)^2\,db
}{
\displaystyle
\int_{b_{\min}}^{b_{\max}} ( I_1(b)^2
+  I_2(b)^2 )\,db
}
\right]^{1/2}.
\ee
compares the radial profiles themselves.

The results of the pairwise comparisons are shown in Tables~\ref{tab2} and~\ref{tab3}. 
The fixed-scale symmetric distances are sizeable, showing that the single-horizon and three-horizon black holes are not pointwise degenerate at the same impact parameter. 
This is expected, because changing the mass parameter shifts the light-ring radius, the critical impact parameter, the disk inner edge and, consequently, the positions of the dominant brightness peaks.

The NRMSE-type diagnostics in Table~\ref{tab3} are larger than the corresponding $L^1$ shape distances, because they give more weight to localized differences in the narrow lensing-ring features. 
They nevertheless support the same qualitative conclusion:
the images are not identical point by point, but their overall morphology is similar once the dominant optical scale is accounted for.

The comparison therefore shows that the similarity seen in Figs.~\ref{pathBimage2} and \ref{pathCimage2} is not an exact equality of the brightness profiles at fixed impact parameter. 
Rather, it is a morphological degeneracy after the overall image scale has been factored out. 
The single-horizon and three-horizon black holes display the same qualitative hierarchy of features: 
a central depression, narrow lensing-ring contributions and a broad direct-emission annulus. 
This supports the conclusion that the number of horizons is not directly encoded in the image morphology; the dominant image features are controlled instead by the 
photon potential and the disk inner edge.

\subsection{Comparison with optical images of other models}

It is instructive to compare the images obtained above with the more familiar cases of Schwarzschild and Reissner-Nordstr\"om black holes, as well as with horizonless compact objects such as boson stars and naked singularities. 
In this subsection we first quantify the comparison with Schwarzschild and Reissner-Nordstr\"om reference images, and then place the results in the broader context of horizonless compact-object images.

The diagnostics introduced in Sec.~5.4 are symmetric, because they compare two solutions of the beyond-Horndeski family without assigning either of them the role of a reference spacetime. 
When comparing with Schwarzschild or Reissner--Nordstr\"om reference images, however, there is a natural reference profile.
We therefore use reference-normalized deviations from Schwarzschild or Reissner--Nordstr\"om images computed with the same disk model and with the matching prescription specified below.

We denote by $I_{\rm BH}(b)$ 
the radial intensity profile of a beyond-Horndeski configuration, and by $I_{\rm GR}(b)$ the corresponding GR reference profile.
At fixed impact parameter we then define
\be
\Delta_{L^1}^{\rm GR}
=
\frac{
\displaystyle
\int_{b_{\min}}^{b_{\max}}
\left|I_{\rm BH}(b)-I_{\rm GR}(b)\right|\,db
}{
\displaystyle
\int_{b_{\min}}^{b_{\max}}
I_{\rm GR}(b)\,db
}.
\ee
This quantity measures the absolute deviation from the GR reference in units of the total reference intensity. 
Since the denominator is fixed by $I_{\rm GR}$, the
diagnostic is intentionally not symmetric: 
it describes how far the beyond-Horndeski image deviates from the reference
GR solution. 

As in Sec.~5.4, one may also remove the overall critical-curve scale before comparing the morphology. 
Defining
\be
\hat b_{\rm BH}=\frac{b}{b_{{\rm crit},{\rm BH}}},
\qquad
\hat b_{\rm GR}=\frac{b}{b_{{\rm crit},{\rm GR}}},
\ee
and
\be
\widetilde I_{\rm BH}(\hat b)
=
I_{\rm BH}(b_{{\rm crit},{\rm BH}}\hat b),
\qquad
\widetilde I_{\rm GR}(\hat b)
=
I_{\rm GR}(b_{{\rm crit},{\rm GR}}\hat b),
\ee
we introduce the reference-normalized shape deviation
\be
\Delta_{L^1}^{\rm GR,shape}
=
\frac{
\displaystyle
\int_{\hat b_{\min}}^{\hat b_{\max}}
\left|
\widetilde I_{\rm BH}(\hat b)-\widetilde I_{\rm GR}(\hat b)
\right|\,d\hat b
}{
\displaystyle
\int_{\hat b_{\min}}^{\hat b_{\max}}
\widetilde I_{\rm GR}(\hat b)\,d\hat b
}.
\ee
The fixed-scale quantity $\Delta_{L^1}^{\rm GR}$ measures deviations at the same impact parameter, while $\Delta_{L^1}^{\rm GR,shape}$ compares the image morphology
after the critical-curve scale has been factored out.

\begin{table}[h]
\centering
\caption{\it Reference-normalized comparison between single-horizon beyond-Horndeski black holes and Schwarzschild black holes with the same ADM mass parameter.} \vspace{0.2cm}
\begin{tabular}{cccccc}
\toprule
Path & $K$ & $\mu_{\rm BH}$ & $\mu_{\rm Sch}$ & $\Delta_{L^1}^{\rm Sch}$ & $\Delta_{L^1}^{\rm Sch,shape}$ \\
\midrule
A & $2.5$ & $2.0$ & $2.0$ & 0.19509 & 0.04008 \\
B & $4.5$ & $1.3$ & $1.3$ & 0.80248 & 0.39733 \\
C & $6.0$ & $1.7$ & $1.7$ & 0.56464 & 0.22451 \\
\bottomrule
\label{tab4}
\end{tabular}
\end{table}

\begin{table}[h]
\centering
\caption{\it Reference-normalized comparison between single-horizon beyond-Horndeski black holes and Schwarzschild black holes with the same horizon radius.\vspace{0.2cm}}
\begin{tabular}{ccccccc}
\toprule
Path & $K$ & $\mu_{\rm BH}$ & $\mu_{\rm Sch}$ & $x_h$ & $\Delta_{L^1}^{\rm Sch}$ & $\Delta_{L^1}^{\rm Sch,shape}$ \\
\midrule
A & $2.5$ & $2.0$ & $1.94613$ & 3.89225 & 0.13107 & 0.04003 \\
B & $4.5$ & $1.3$ & $0.99685$ & 1.99369 & 0.64752 & 0.39134 \\
C & $6.0$ & $1.7$ & $1.49110$ & 2.98219 & 0.37733 & 0.23747 \\
\bottomrule
\label{tab5}
\end{tabular}
\end{table}

Tables~\ref{tab4} and~\ref{tab5} show that the degree of agreement with Schwarzschild depends strongly on the matching prescription. 
At equal ADM mass parameter, the path~A black hole is quantitatively close to Schwarzschild, especially after rescaling by $b_{\rm crit}$, whereas the path~B and path~C examples show larger deviations. 
Matching the horizon radius instead reduces the fixed-scale deviations, most clearly in path~C, but does not remove all differences in the radial profiles. 
Thus part of the deviation from Schwarzschild is an overall image-scale effect, while the remaining difference reflects changes in the transfer function and in the relative location of the disk image features.

In the Schwarzschild black hole, the thin disk produces a central dark region, a narrow lensing or photon-ring structure close to the critical curve, and a broader direct image of the disk~\cite{Gralla:2019xty}. 
The qualitative origin of these features is well understood:
the unstable photon sphere determines
the strong-deflection region, while the ISCO fixes the inner edge of the emitting disk. 
Several of our single-horizon black-hole images with an ISCO show precisely this pattern. 
For example, the large-mass configurations in paths A, B and C display a dark central region, narrow higher-order ring contributions, and a broad outer annulus. 
In this sense, their optical appearance is Schwarzschild-like even though the underlying metric is not Schwarzschild. 
The quantitative comparison in Tables~\ref{tab4} and~\ref{tab5} refines this statement: 
the images may share the same hierarchy of features while still showing sizeable radial-profile deviations from Schwarzschild for some mass paths.

The analogy with Reissner--Nordstr\"om black holes is also instructive, especially for the two-horizon and multi-horizon branches. 
For the RN comparison, the charge parameter $Q_{\rm RN}$ is used only as a parameter of the reference metric. 
It is fixed by the matching prescription and should not be interpreted as an electromagnetic charge of the beyond-Horndeski spacetime. 
In Table~\ref{tab6} we choose the RN reference to have the same ADM mass parameter and the same outer horizon radius as the corresponding beyond-Horndeski solution.

\begin{table}[h]
\centering
\caption{\it Reference-normalized comparison between beyond-Horndeski black holes and Reissner--Nordstr\"om reference black holes. 
The RN reference is chosen to have the same ADM mass parameter and the same outer horizon radius as the beyond-Horndeski solution. 
The parameter $Q_{\rm RN}$ is therefore only a parameter of the reference metric and should not be interpreted as a physical charge of the beyond-Horndeski solution.
\vspace{0.2cm}}
\begin{tabular}{cccccccc}
\toprule
Path & $K$ & $\mu_{\rm BH}$ & $N_{\rm H}$ & $x_{h+}$ & $Q_{\rm RN}$ & $\Delta_{L^1}^{\rm RN}$ & $\Delta_{L^1}^{\rm RN,shape}$ \\
\midrule
B & $4.5$ & $1.205$ & 3 & 1.54048 & 1.15736 & 0.74515 & 0.24901 \\
C & $6.0$ & $1.414$ & 2 & 2.05717 & 1.25926 & 0.68774 & 0.20866 \\
C & $6.0$ & $1.5$ & 3 & 2.39354 & 1.20482 & 0.52217 & 0.13410 \\
\bottomrule
\label{tab6}
\end{tabular}
\end{table}

The RN comparison shows the same general pattern as the Schwarzschild comparison:
the shape deviations are smaller than the fixed-scale deviations. 
This indicates that part of the difference is due to an overall shift of the optical scale. 
Nevertheless, the remaining shape deviations are not negligible. 
Thus the multi-horizon beyond-Horndeski black holes are not simply equivalent to RN black holes at the level of the thin-disk transfer function, even when the ADM mass parameter and the outer horizon radius are matched.

A RN black hole may contain an inner horizon in addition to the outer event horizon, but the image seen by a distant observer is mainly determined by the outer photon potential
and by the disk emission region. 
The inner horizon does not directly participate in the
transfer function of photons that reach infinity.
This is similar to what we find for the multi-horizon black holes in our model. 
In both paths B and C, a three-horizon black hole can have an optical appearance close to that of a single-horizon black hole, provided that the outer light ring and the disk inner edge are similar. 
Therefore, the number of horizons is not by itself an observable image diagnostic in this thin-disk model.

A more important distinction arises when the disk inner edge is not an ISCO but a static sphere.  
In such cases the image scale can be much smaller than in the usual Schwarzschild-like ISCO images.  
The central depression is then strongly affected by the location of the static sphere and by the transfer function.
This behavior has no direct analogue in the standard Schwarzschild thin-disk picture, where the disk inner edge is usually placed at the ISCO. 
It is one of the characteristic signatures of the present family of solutions.

Our horizonless configurations should also be compared with boson stars\cite{Rosa:2022tfv} and naked singularities \cite{Gyulchev:2019tvk,Shaikh:2019hbm,Joshi:2020tlq,Dey:2020bgo,Deliyski:2026fav} studied in the literature.  
A common feature from those systems is that a central dark region in the image is not necessarily an event-horizon shadow.  
Boson stars, naked singularities, and other horizonless compact objects may produce shadow-like depressions when the emission has an inner cutoff or when the transfer function prevents photons from sampling the central emitting region efficiently.  
The naked singularities in our model show the same phenomenon.  
In paths A, B and C, the central dark region is caused by the disk inner edge and by lensing, rather than by photon capture at a horizon.  
Thus the dark region should be interpreted as a shadow-like feature, not as a genuine black-hole shadow.

There are nevertheless differences from previously studied naked singularity images.  
In the present solution, some naked singularities possess sufficiently strong deflection to generate an additional inner bright ring.  
This ring is not associated with a logarithmic divergence of the deflection angle at an unstable light ring.  
Instead, it is produced by an additional lensed image of the disk inner edge.  
This provides a useful way to distinguish different horizonless geometries: two objects may both lack horizons, but their transfer functions can have different branch structures and hence different ring patterns.

Overall, the comparison shows that the optical image is determined less by the global classification of the spacetime and more by a small set of 
geometric and disk quantities, such as the outer unstable light ring, the critical impact parameter, the disk inner edge, and the structure of the transfer function. 
A Schwarzschild-like image does not uniquely imply a Schwarzschild black hole, and a shadow-like central depression does not uniquely imply an event horizon.  
Conversely, genuinely different horizon structures, including multi-horizon black holes, can be morphologically close in optical appearance if their exterior photon potentials and disk inner edges are close, although the quantitative radial-profile diagnostics still reveal differences at fixed impact parameter.

\section{Conclusions}

In this work we have studied the geodesic structure and optical appearance of compact objects with primary scalar hair in shift- and parity-symmetric beyond Horndeski gravity.
Focusing on the analytic $n/s=5/2$ solution, we classified the spacetime branches at fixed theory parameter $K$ as the dimensionless mass parameter $\mu$ is varied. 
Depending on $K$, the same family of solutions interpolates between timelike naked singularities, regular solitons, regular black holes, Reissner--Nordstr\"om-like black holes, multi-horizon black holes, and Schwarzschild-like single-horizon black holes. 
This solution therefore provides a single analytic family in which qualitatively different compact-object branches can be compared at fixed theory parameters.

We then analyzed the null and timelike geodesic structure of these spacetimes. 
The number and location of unstable light rings were found to depend sensitively on the structure of the photon potential and do not follow simply from the number of horizons. 
In particular, some horizonless configurations possess light rings, while some single-horizon black holes admit more than one light ring. 
We also determined the regions in parameter space where the inner scale of a thin disk is set either by an ISCO or by a static sphere. 
This shows that the transition between different compact object branches is reflected not only in the horizon structure, but also in the organization of circular photon and particle orbits.

Finally, we computed optical images produced by a geometrically and optically thin disk viewed face-on by a distant observer. 
The images show that the number of horizons is not directly encoded in the observed brightness distribution. 
Horizonless objects can exhibit shadow-like central depressions when the disk has an inner edge and the transfer function prevents photons from sampling the central region.
Conversely, multi-horizon black holes can closely resemble ordinary single-horizon black holes when their exterior light-ring structure and disk inner edge are similar. 
The optical appearance is therefore governed mainly by the exterior photon potential and the location of the disk inner edge, while the deeper horizon structure leaves only an indirect imprint.

The quantitative diagnostics introduced in Sec.~5.4 refine this conclusion. 
At fixed impact parameter, the radial intensity profiles of the single-horizon and three-horizon black holes are not pointwise degenerate, because the light-ring radius, critical impact parameter and disk inner edge shift along the mass path.
However, after rescaling by the critical impact parameter, the shape diagnostics are significantly smaller. 
This shows that the degeneracy is primarily morphological: different horizon structures can produce the same hierarchy of image features, namely a central depression, narrow lensing-ring contributions and a broad direct-emission annulus.

The comparison with Schwarzschild and Reissner--Nordstr\"om reference images further supports this interpretation. 
The agreement with Schwarzschild depends on the matching prescription: 
matching the horizon radius can reduce the fixed-scale deviation, while shape-rescaled diagnostics isolate the remaining morphological differences. 
Similarly, the Reissner--Nordstr\"om comparison shows that multi-horizon beyond-Horndeski black holes can share some scale-free image features with two-horizon reference geometries, but are not equivalent to them at the level of the thin-disk transfer function.

As an outlook, it would be interesting to extend this analysis in several directions. 
A natural next step is to consider inclined disks, where Doppler boosting and image asymmetry can provide additional information beyond the face-on brightness profiles studied here. 
One may also replace the simple phenomenological emissivity by more realistic disk models, include optically thin extended emission, or perform full radiative-transfer calculations. 
Finally, rotating generalizations of these solutions would allow a closer comparison with Kerr black holes and with horizon-scale observations of compact objects such as those obtained by the Event Horizon Telescope.
Such extensions would make it possible to assess more directly whether the degeneracies found here between horizonless objects, multi-horizon black holes and single-horizon black holes can be broken by realistic horizon-scale observations.

\section*{Acknowledgments}

We gratefully acknowledge communications with A.~Bakopoulos for pointing us to the interesting analytic solution studied here.
This work is supported by the National Natural Science Foundation of China (NSFC) under Grant Nos.~12565010 and 12205123.

\end{document}